\documentclass[conference]{IEEEtran}

\ifCLASSOPTIONcompsoc
  \usepackage[nocompress]{cite}
\else
  \usepackage{cite}
\fi

\usepackage{tikz}
\usepackage{amsmath}

\usepackage[utf8]{inputenc}
\usepackage{xcolor}
\usepackage{graphicx}
\usepackage{hyperref}
\usepackage{amssymb,amsfonts}
\usepackage{listings}
\usepackage{array}

\usepackage{booktabs, multirow} % for borders and merged ranges
\usepackage{soul}% for underlines
\usepackage{colortbl} % for cell colors
\usepackage{changepage,threeparttable}
\usepackage{tabularx}
\usepackage[normalem]{ulem} % for \sout

\usepackage[font=footnotesize]{subcaption}
\usepackage[font=footnotesize]{caption}

\usepackage{algorithm}
\usepackage{algpseudocode}

\usepackage{fontawesome5}
\usepackage{wasysym}

\usepackage{tcolorbox}

\newcommand{\ie}[0] {i.e., }
\newcommand{\tableRef}[1] {Table~\ref{#1}}
\newcommand{\fig}[1] {Fig.~\ref{#1}}
\newcommand{\figRef}[1] {Fig.~\ref{#1}}
\newcommand{\secRef}[1] {Sec.~\ref{#1}}
\newcommand{\appRef}[1] {Appendix ~\ref{#1}}

\newcommand{\para}[1] {\smallskip \noindent {\em #1:}}

\newcommand{\parabf}[1] {\smallskip \noindent {\bf #1:}}
\newcommand{\equifaxAttackStage}[0] {St\xspace}

\newcommand{\actionModule}[0]{action translator}
\newcommand{\ActionModule}[0]{Action translator}

\newcommand{\telemetryModule}[0]{observability module}
\newcommand{\TelemetryModule}[0]{Observability module}

\newcommand{\equifaxEnvShort}[0]{Equifax-inspired environment}
\newcommand{\cpEnvShort}[0]{Colonial Pipeline-inspired environment}

\newcounter{finding}
\makeatletter
{
    \refstepcounter{finding}%
    \protected@edef\@currentlabelname{#1}% addition here
    \noindent
    {
        \begin{tcolorbox}
        {
            \textbf{Finding \thefinding}: #1
        }
        \end{tcolorbox}
    }
}
\makeatother

\newcounter{takeaway}
\makeatletter
{
    \refstepcounter{takeaway}%
    \protected@edef\@currentlabelname{#1}% addition here
    \noindent
    {
        \noindent
        \begin{tcolorbox}
        {
            \textbf{Takeaway \thetakeaway}: #1
        }
        \end{tcolorbox}
    }
}
\makeatother

\newcounter{challenge}
\makeatletter
{
    \refstepcounter{challenge}%
    \protected@edef\@currentlabelname{#1}% addition here
    \vspace{0.2cm}
    \noindent
    {
        \vspace{1mm}
        \noindent
        \begin{tcolorbox}
        {
            \textbf{Challenge \thechallenge}: #1
        }
        \end{tcolorbox}
    }
    {\vspace{.2mm}}
}
\makeatother

\newcounter{packednmbr} % Define the counter if not already defined

\newenvironment{packeditemize}{
  \begin{list}{$\bullet$}{
    \setlength{\itemsep}{1pt}
    \addtolength{\labelwidth}{12pt}
    \setlength{\leftmargin}{\labelwidth}
    \setlength{\listparindent}{\parindent}
    \setlength{\parsep}{1pt}
    \setlength{\topsep}{1pt}
  }
}{
  \end{list}
}

\newcommand{\tightcaption}[1]{
    \vspace{-0.2cm}
    \caption{#1}
    \vspace{-0.2cm}
}

\newcommand*\emptycirc[1][0.75ex]{\tikz\draw (0,0) circle (#1);} 

\newcommand*\fullcirc[1][0.75ex]{\tikz\fill (0,0) circle (#1);} 

\definecolor{codegreen}{rgb}{0,0.6,0}
\definecolor{codegray}{rgb}{0.5,0.5,0.5}
\definecolor{codepurple}{rgb}{0.58,0,0.82}
\definecolor{backcolour}{rgb}{0.97,0.97,0.97}

\definecolor{white}{rgb}{1,1,1}

\definecolor{githubOrange}{HTML}{c9510c}
\definecolor{githubPurple}{HTML}{6e5494}
\definecolor{githubGreen}{HTML}{6cc644}
\definecolor{githubBlue}{HTML}{4078c0}
\definecolor{githubGrey}{HTML}{f5f5f5}

\lstdefinestyle{mystyle}{
    language=Python,
    backgroundcolor=\color{white},   
    commentstyle=\color{codegreen},
    keywordstyle=\color{githubOrange},
    numberstyle=\color{githubBlue},
    stringstyle=\color{githubBlue},
    basicstyle=\ttfamily\footnotesize,
    breakatwhitespace=false,
    breaklines=true,                 
    captionpos=b,                    
    keepspaces=true,                 
    numbers=none,                    
    numbersep=5pt,                  
    showspaces=false,                
    showstringspaces=false,
    showtabs=false, 
    tabsize=2,
}

\begin{document}

\newcommand{\platform}{{Perry}}

\title{\platform: A High-level Framework for \\  Accelerating Cyber Deception Experimentation}

\author{
    \IEEEauthorblockN{Brian Singer}
    \IEEEauthorblockA{\em Carnegie Mellon University}
    \and
    \IEEEauthorblockN{Yusuf Saquib}
    \IEEEauthorblockA{\em Carnegie Mellon University}
    \and
    \IEEEauthorblockN{Lujo Bauer}
    \IEEEauthorblockA{\em Carnegie Mellon University}
    \and
    \IEEEauthorblockN{Vyas Sekar}
    \IEEEauthorblockA{\em Carnegie Mellon University}
}

\maketitle

\begin{abstract}
	Cyber deception aims to distract, delay, and detect network attackers with fake assets such as honeypots, decoy credentials, or decoy files. 
However, today, it is difficult for operators to experiment, explore, and evaluate deception approaches.
Existing tools and platforms have non-portable and complex implementations that are difficult to modify and extend. 
We address this pain point by introducing Perry, a high-level framework that accelerates the design and exploration of deception what-if scenarios.
Perry has two components: a high-level abstraction layer for security operators to specify attackers and deception strategies and an experimentation module to run these attackers and defenders in realistic emulated networks.
To translate these high-level specifications into low-level primitives, we design four key modules in Perry: 1) an action planner that translates high-level actions into low-level implementations, 2) an observability module to translate low-level telemetry into high-level observations, 3) an environment state service that enables environment agnostic strategies, and 4) an attack graph service to reason how attackers could explore an environment.
We illustrate that Perry’s abstractions reduce the implementation effort across a wide variety of deception defenses, attackers, and environments.
We demonstrate the value of Perry by emulating 55 unique deception what-if scenarios, illustrating how these experiments enable operators to shed light on subtle tradeoffs.

\end{abstract}

\section{Introduction}
% Deception is promising
Cyber-deception-based defenses seek to deceive an attacker with fake resources, information, or critical assets.
These could include a decoy host (e.g., a honeypot)~\cite{provos2004_honeypot, pa2016iotpot, moore2016malwareHoneypot}, decoy credentials~\cite{amnesia_honey_db, han2018_deception_rol}, decoy files~\cite{han2018_deception_rol}, or other system assets. 
Cyber deception holds promise to distract attackers~\cite{ferguson2021_deception_psychology, han2018_deception_rol}, detect attacks in progress~\cite{amnesia_honey_db, heckman2015_deception_book}, and delay attackers~\cite{ferguson2021_deception_psychology}.
With emerging AI-based attackers~\cite{ deng2024pentestgpt, rodriguez2025framework}, cyber deception may become a critical part of the defense arsenal~\cite{kouremetis2025occult}.

In order to inform their future  security posture,  security operators need the ability to  
run a broad spectrum of \emph{deception what-if scenarios} relevant to their environment.
For example: 
Would deploying many decoy files to a network distract attackers? 
Where are the optimal locations to deploy decoy credentials in a network? 
From these experiments, security operators could collect quantitative measurements on cost (e.g., number of decoys used) and effectiveness (e.g., data exfiltrated, hosts infected).  

Unfortunately, running such experiments with existing tools such as Caldera and Elastic Security~\cite{caldera, mirage, cyborg, holm2022lore, ibmSecurityQRadar, elasticsecurity} entails significant operator effort and complexity.
We find that current tools for specifying attacker (e.g., execute  exploits, command infected computers) and deception (e.g., detect fake credential usage) strategies operate at a very low level of abstraction.
These tools rely on low-level mechanisms such as command-line interfaces, manual configuration files, and raw network log parsing, making them complex and cumbersome to use.
The lack of abstraction and modularity also means that changes to attacker and defender implementations entail complex and distributed changes.
Thus, operators cannot easily take existing implementations of  strategies and quickly port it to work in their own environments and experiments. 
Taken together, this means that operators are stymied in their efforts to run deception-related experiments.

\begin{figure}
    \centering
    \includegraphics[width=1.0\linewidth]{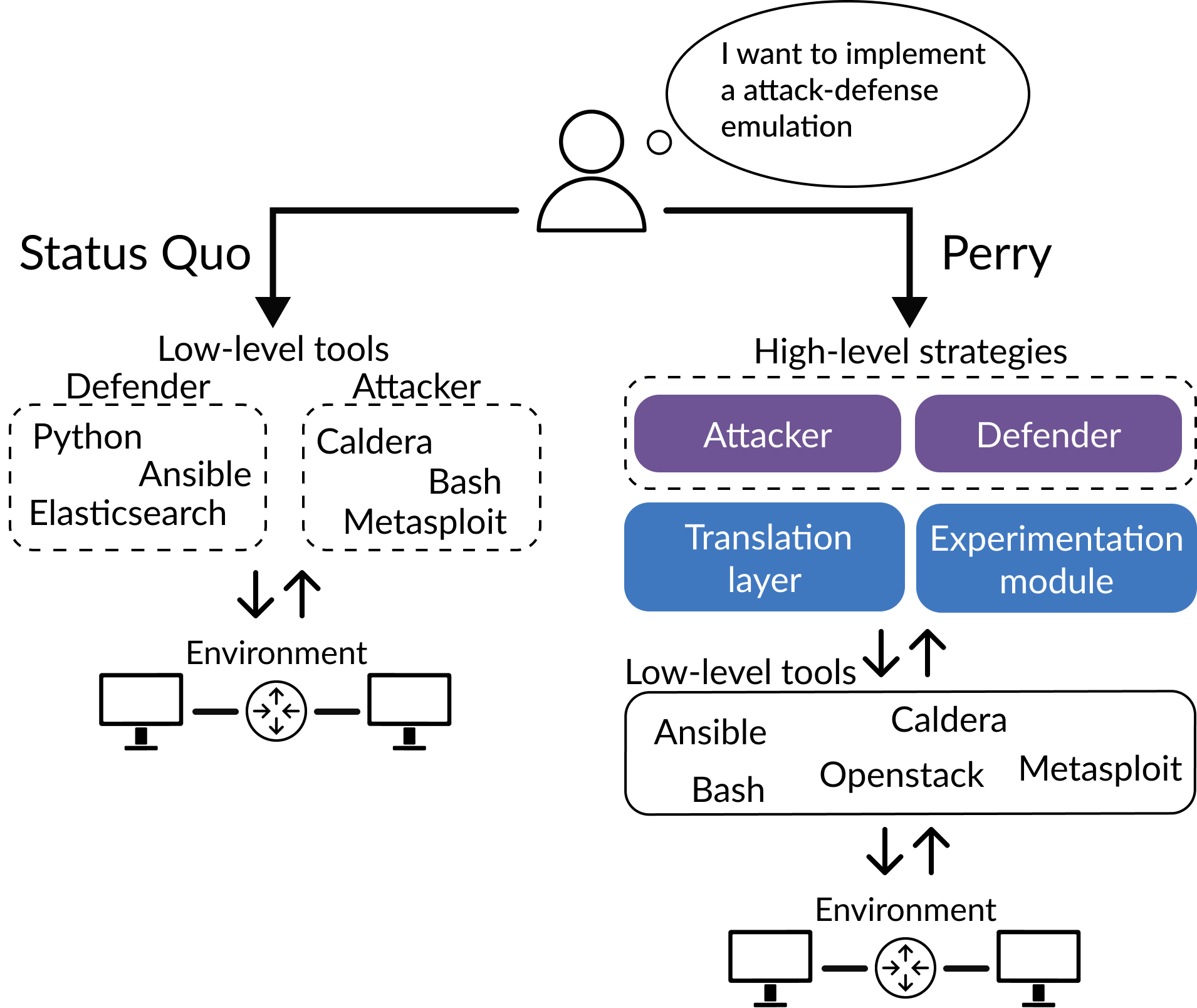}
    \caption{Currently, operators have to use low-level tools to build deception what-if scenarios.
    We accelerate the process of implementing deception what-if scenarios by introducing Perry.
    With Perry, operators can quickly program deception what-if scenarios at a high-level, and Perry will translate scenarios into their low-level primitives.}
    \label{fig:intro_perry}
\end{figure}

To address this challenge, we present Perry, a high-level framework for accelerating cyber deception experimentation.
Perry has two components: (1) a high-level abstraction layer for security operators to specify attackers and deception defenders in a portable environment-agnostic manner and (2) an experimentation module to run these attackers and defenders in realistic emulated networks (\figRef{fig:intro_perry}).
 Operators use a high-level and environment-agnostic state machine abstraction to specify attack and deception strategies, which execute {\em high-level actions} based on {\em high-level observations} of the environment (\secRef{sec:system_overview}).   
Because these specifications are environment agnostic, they are naturally portable from one setting to another; for example,  operators can take  open-source contributions of existing attacker and defender strategies (either from Perry developers and the broader community) and run them seamlessly in their own emulated environments  (with comparable telemetry and actuation capabilities)  with low effort. 

Perry uses four key modules to translate high-level attacker and deception strategies into low-level primitives: 1) an \actionModule\ to specify attack and defense actions at a high level (e.g., deploy a decoy), 2) an event-driven \telemetryModule\ that emits high-level observations (e.g., host is suspicious), 3) an environment state service to enable environment agnostic strategies, and 4) an attack graph service to reason how an attacker can explore an environment.

We show that  Perry's abstractions can reduce the implementation effort across a wide variety of defense and attack strategies (see \secRef{sec:eval_rapid_exploration}).
We validate that there is no  effort needed to port attacker and deception strategies across environments  due to Perry's environment-agnostic and portable specifications. We quantify  the  reduction in lines of code (LOC) by 17--27$\times$ for attacker implementations and 7--14$\times$ for deception implementations.
We also show how to optionally use LLM code generation to create correct implementations of deception strategies in Perry.
For instance, in \secRef{sec:eval_rapid_exploration}, we evaluate two attackers and two deception approaches across five environments.
With Perry, users only need to write each attack and deception strategy once, and these can be applied to all environments without additional effort. 
In contrast, previous frameworks require users to manually re-implement each attack and defense strategy for each environment.
In \secRef{sec:eval_rapid_exploration}, we also measure the LOC of deception approaches and attackers with and without using Perry’s abstractions.

This reduction in effort can enable operators to rapidly 1) explore a variety of deception strategies against different attackers in different environments (see \secRef{sec:eval_rapid_exploration}), 2) implement variations of strategies  (see \secRef{sec:eval_perry_mutations}), and 3) evaluate extensions to consider new defense and attacker capabilities (see \secRef{sec:eval_perry_extensions}).
Using Perry, we explore diverse scenarios and shed light on the efficacy of deceptive defenses (see \secRef{sec:eval_findings}). For instance,
\begin{packeditemize}
  \item Fine-grained and host telemetry that use system calls to detect attackers interacting with decoy resources can reduce data exfiltration by 40 to 100\%.
  \item Simple extensions to deception strategies can have significant impacts on efficacy. In one case, a defender purposefully leaking a deceptive network topology to attackers slowed an attacker down by 3.2x.
  \item Emerging LLM-based autonomous attackers (e.g., \cite{cyberseceval3}) can be slowed down using decoy-based deception.  
  We show how these LLM attackers can spend 88-92\% of their commands interacting with decoy hosts.
\item The efficacy of deception strategies can vary widely across different attackers and environments, highlighting the importance of rapid exploration. For instance, a layered deception strategy can delay attackers anywhere from 3\% to 49\% depending on the attack strategy.
\end{packeditemize}

\noindent \textbf{Contributions and Roadmap:} In summary, this paper makes the following contributions:
\begin{packeditemize}
    \item We introduce a novel high-level and environment agnostic abstraction for expressing attackers and deception defenders: a state machine that executes high-level actions based on high-level observations of the environment (\secRef{sec:system_overview}).
    \item We present a concrete design and implementation to translate  high-level intentions into their lower level  primitives (\secRef{sec:detailed_system}--\secRef{sec:implementation}).
    \item We demonstrate the value of Perry by illustrating how it can enable operators to rapidly explore a wide variety of deception defenses, attackers, and environments, shedding light on the tradeoffs of deception defenses(\secRef{sec:eval}).  
\end{packeditemize}

\textbf{Reproducible Research} Perry is open-source and will be publicly available to the research community.\footnote{The code is available at \hyperlink{https://github.com/bsinger98/Perry}{https://github.com/bsinger98/Perry}.}

\section{Background and Motivation} \label{sec:motivation}
In this section, we discuss the importance for operators to quickly and systematically evaluate deception approaches against candidate attackers.
We begin by making the case for what-if experiments in cyber deception (\secRef{sec:motivation_case_for_what_if}).
Then, we use an illustrative example to show that existing tools entail significant complexity and are difficult to extend (\secRef{sec:motivation_limitations}). 

\subsection{The case for what-if experiments in deception }
\label{sec:motivation_case_for_what_if}
To illustrate the need for the rapid exploration of what-if experiments, we examine a real world incident, the Equifax data breach, as a motivating  example.

\begin{figure}[tb]
    \centering
    \includegraphics[width=0.40\textwidth]{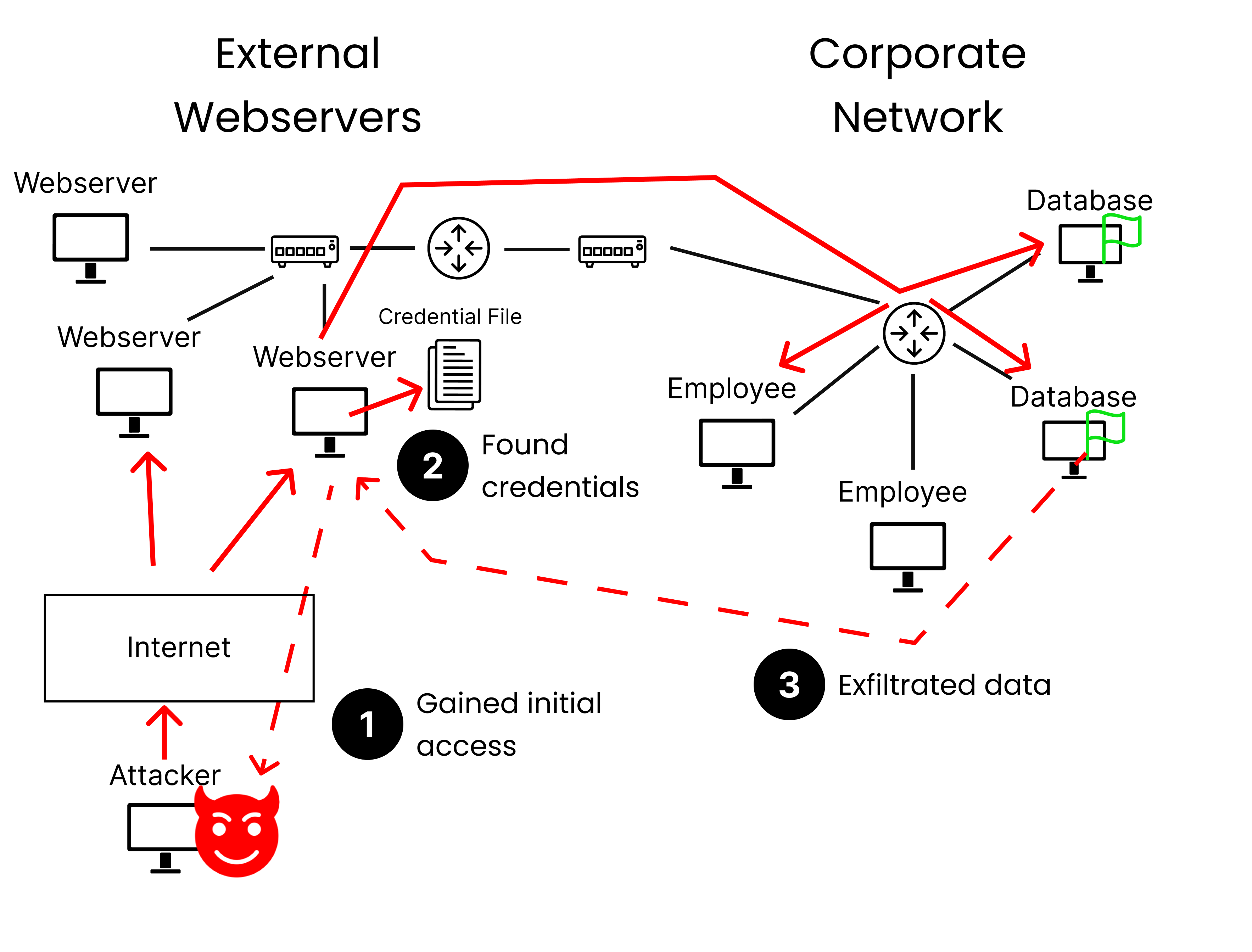}
    \tightcaption{The attack that led to the Equifax data breach had three stages.
    First, the attackers gained initial access by compromising external web servers.
    Next, they found credentials
    to databases on a web server.
    Finally, the attackers exfiltrated critical user data from the databases.} 
    \label{fig:equifax_scenario}
\end{figure}

\begin{figure*}[tb]
    \centering
    \includegraphics[width=.8\textwidth]{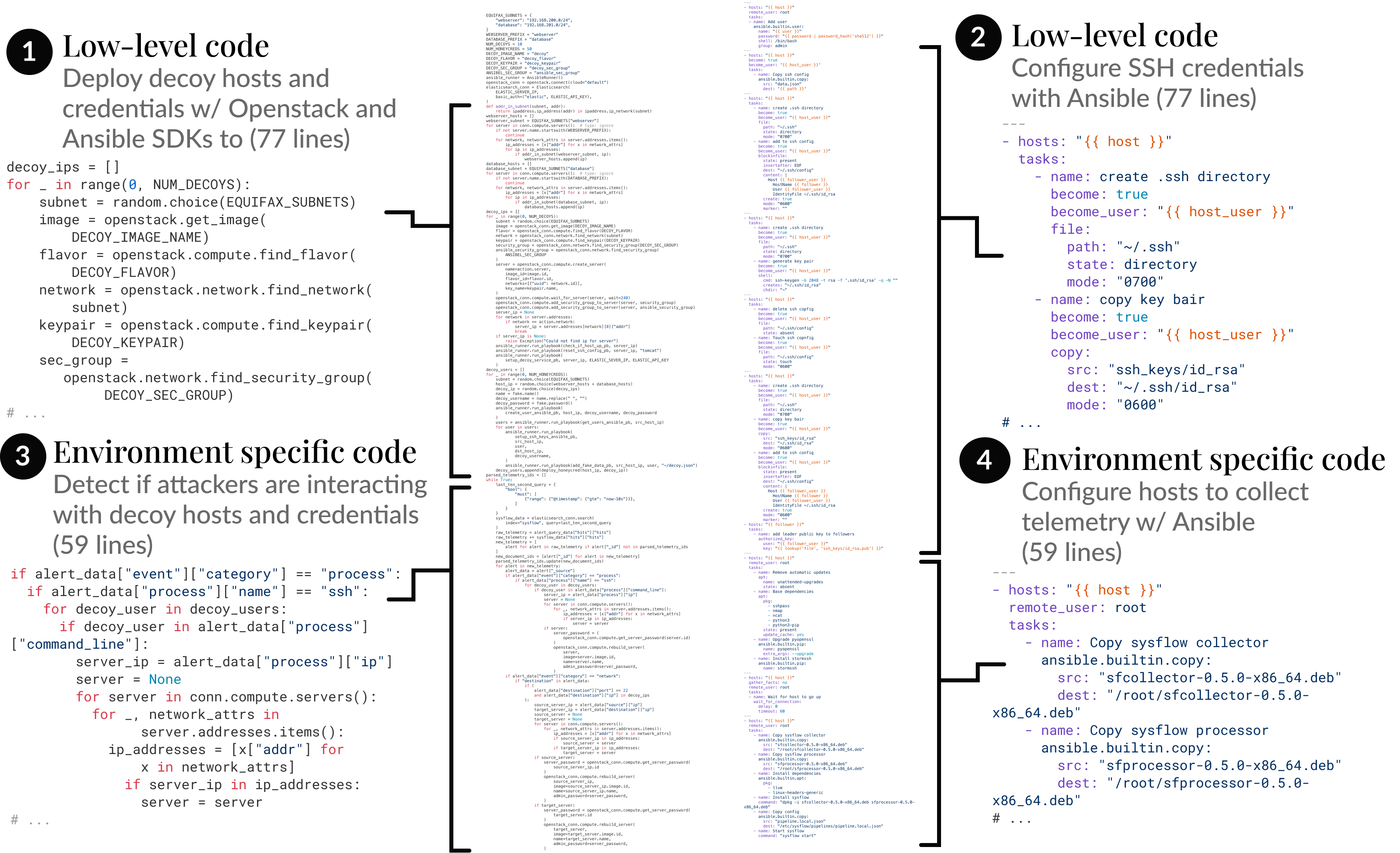}
    \tightcaption{Code snippets of the stateful deception strategy implemented using Elasticsearch and Python scripts. 
    The implementation requires 467 lines of low-level code that
    use the low-level Openstack, Elasticsearch, and Ansible
    SDKs.
    In addition, the implementation has code intertwined with the environment to detect interactions to decoy hosts and resources.}
    \label{fig:defender_before_snippet}
\end{figure*}

In 2017, attackers stole an estimated 143 million private American records from Equifax~\cite{equifax_report}. 
Equifax was forced to pay over \$500 million in legal penalties as a direct result of the attack~\cite{equifax_ftc_settlment}.  
As shown in \figRef{fig:equifax_scenario}, the attack that led to
the Equifax breach had three stages: 
(1) The attackers gained initial access to the Equifax network through
a known vulnerability on their web servers. (2) The attackers
discovered an unencrypted file on the web servers containing
credentials to databases. (3) The attackers
utilized the credentials to access and exfiltrate data from the
databases~\cite{equifax_report}. 

Following this example, if and how, could deception techniques have stopped the attack:
Would some deception strategies have  higher efficacy than others?
To shed light on these questions, operators may want to explore what-if scenarios: 

\noindent \textbf{Example 1.} The operator may want to quickly explore the efficacy of a wide variety of deception capabilities (e.g., decoy hosts~\cite{pa2016iotpot}, fake credentials~\cite{amnesia_honey_db}, or fake software patches~\cite{araujo2020_honeypatch}) against many variants of this attacker.

\noindent \textbf{Example 2.} The  operator may want to understand the benefits of adding better telemetry with the above deception algorithms,  enabling the defense to react to attackers actions.
For example, an operator may want to detect and react to attackers interacting with deceptive fake credentials the defender added.

\subsection{Requirements and limitations of prior work} \label{sec:motivation_limitations}

In order for operators to quickly implement deception what-if experiments, we require a platform that is:
\begin{packeditemize}

\item {\em Expressive}: A platform should be capable of modeling  diverse scenarios spanning many types of deception approaches, attack strategies, and environments. 

\item {\em Low effort}: We want to quickly implement and evaluate diverse scenarios; e.g., reuse components and code as much possible.

\item {\em Extensible}: As new deception approaches, system capabilities, and attack strategies
appear; the platform should be extensible to evaluate these new techniques.  

\end{packeditemize}

At one extreme, we can consider simple agent-based simulators that are low effort~\cite{milani2020attackGraph, kiekintveld2015honeypotGameTheory, cranford2018stackelberg, msft:cyberbattlesim}.
These do not emulate real networks, attackers, or defenders and lack the realism and expressiveness for real-world scenarios.
We will later see in \secRef{sec:eval_rapid_exploration} how subtle implementation details can greatly impact the efficacy of deception.

At the other extreme, we have a range of tools (e.g., Caldera, Metasploit, Snort, Elasticsearch~\cite{snort, metasploit, caldera, elasticsearch}) that can realistically emulate real attacks and defenses.
Let us consider implementing Example 2 with these low-level tools. 
We can implement the attacker with
Caldera~\cite{caldera}, a widely used tool in deception evaluations, such as CybORG~\cite{cyborg}, Mirage~\cite{mirage}, and CyGil~\cite{cygil}.
We can implement defenders by combining SIEM tools like Elasticsearch~\cite{elasticsecurity} with custom Python scripts.
To deploy the environment (i.e., the network, hosts, and critical assets) we can use an OpenStack-like virtual network~\cite{openstack}.

Specifically, we consider a stateful defender (Example 2) that 1) deploys decoy hosts and credentials to the network, 2) detects hosts interacting with the decoy hosts and using the decoy credentials, and 3) restores any hosts that interact with decoy resources.
Similarly, we create an attacker that mimics the real-world Equifax attacker~\cite{equifax_report} by: 1) gaining initial access to the network by infecting vulnerable web servers, 2) searching and using credentials to infect database hosts, and 3) exfiltrating data from each database host. \figRef{fig:defender_before_snippet} shows code snippets of the defender implementation. 

Next, we highlight three issues with such   implementations.\footnote{
Our goal is to illustrate problems broadly with available tooling, not at these specific tools.
For instance, the problems we highlight with  Caldera also exist in other attacker emulation tools~\cite{metasploit, holm2022lore, enoch2020harmer} and 
the pain points of Elasticsearch with Python are similar to other defense tools~\cite{snort, ibmSecurityQRadar, splunkSOAR}.) }

\begin{packeditemize}

\item{\bf Low-level and cumbersome code:}
Existing tools offer low-level SDKs resulting in cumbersome low-level code.
For instance, the stateful defender implementation requires 272 lines of low-level code.
Deploying decoy hosts and credentials requires 213 lines of code using the Ansible and Openstack SDKs.
The remaining 59 lines of code use the Elasticsearch API to detect attackers interacting with decoy hosts and credentials.

Attacker implementation tools also have low-level SDKs.
For example, to infect the web servers, we have to add low-level code for each type of network scan and exploit script.
In another case, to find credentials, we had to: 1) implement bash scripts to output credentials, 2) implement parsers to interpret the output of the bash commands, and 3) implement the attacker strategy to call the bash script.

\item{\bf Non-portable code:} Existing tools force the code to be  tightly coupled with the environment, making it difficult to port attackers and defenders to a new environment.
For instance, the attacker implementation exfiltrates data.
In the Equifax environment, this requires (1) choosing a correct path of two hosts to exfiltrate the data on, and (2) choosing the correct exfiltration protocol between the hosts (e.g., the web server communicates with the database instance with \texttt{SSH} while the attacker and web server communicate with \texttt{HTTPS}).
If the operator's environment has a different path or different protocols, the attacker implementation will fail.

\item {\bf Modifications require distributed changes:}
As a consequence of complex low-level code, existing tools have complex dependencies between strategies, capabilities, or telemetry, making it difficult to extend to evaluate new deception approaches and attackers.
Our implementations became tightly coupled to their capabilities and telemetry, making them difficult to extend.
For instance, we may want to extend our deception to add decoy files to decoy hosts.
But adding decoy files requires many distributed changes to the implementation.
Specifically, in \figRef{fig:defender_before_snippet}, we have to: (1) extend the initialization function to maintain lists of real and decoy file, (2) extend the detection rules to detect if fake data is being exfiltrated, (3) and (4) implement Ansible Playbooks to deploy the decoy files.

\end{packeditemize}

\section{\platform\ System Overview}
\label{sec:system_overview}
In this section, we give an overview of Perry.
Perry introduces a high-level abstraction layer for security operators to specify attackers and deception defenders as well as an experimentation module to emulate these attackers and defenders in realistic network topologies.

\begin{figure}[tb]
    \centering
    \includegraphics[width=0.45\textwidth]{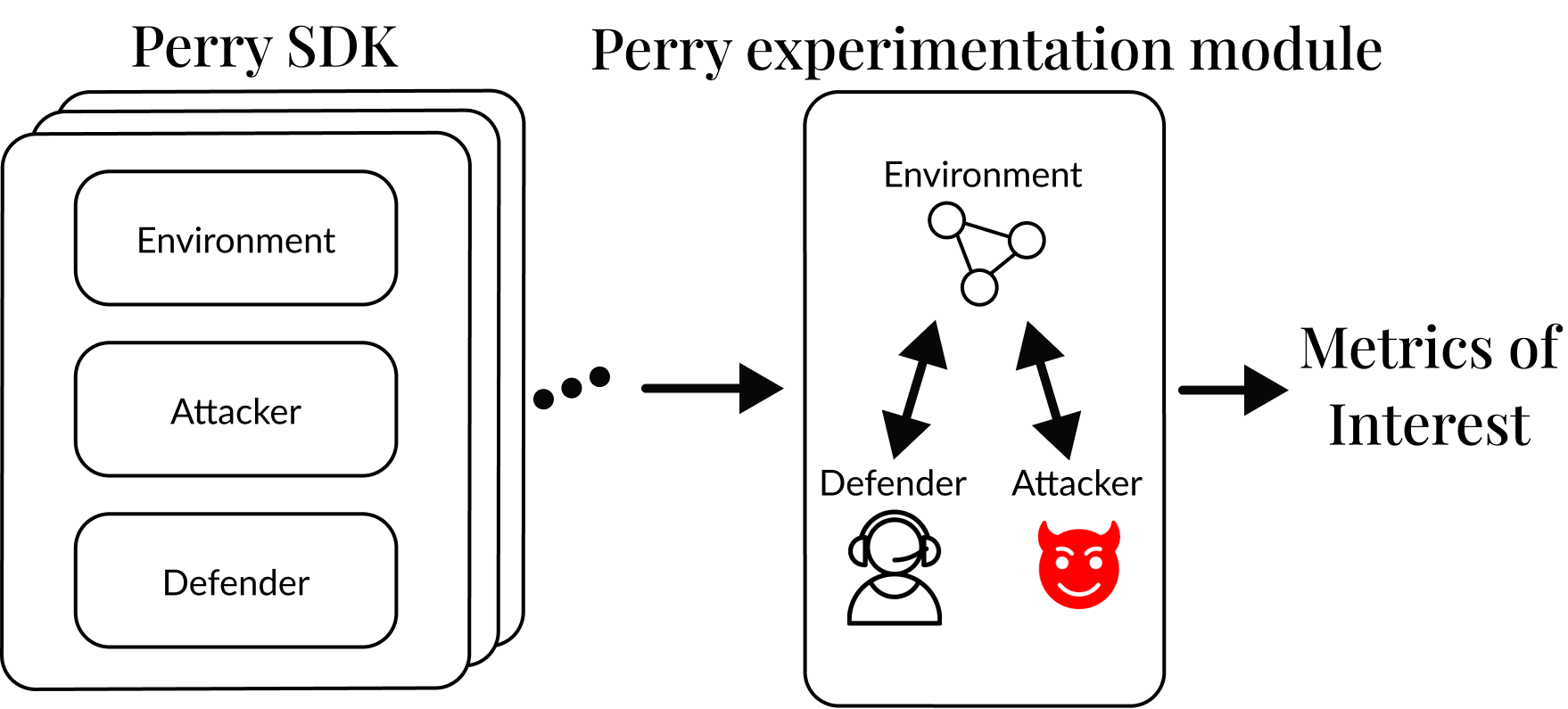}
    \tightcaption{A system overview of Perry. Operators use the Perry SDK to specify the environment, attacker, and defender to execute. Perry then executes each scenario and outputs metrics of interest.}
    \label{fig:perry_overview}
\end{figure}

The inputs to Perry are the environment, attacker, and defender, shown in \figRef{fig:perry_overview}.
Perry first instantiates the environment on a virtualized network and  executes the defender and attacker logic expressed using its high-level abstraction. 
Perry collects metrics such as defender resources used (e.g., number of decoys used) and attacker success metrics (e.g., data exfiltrated) to enable  quantitative analytics (\secRef{sec:eval}).

\begin{figure}[tb]
    \centering
    \includegraphics[width=0.35\textwidth]{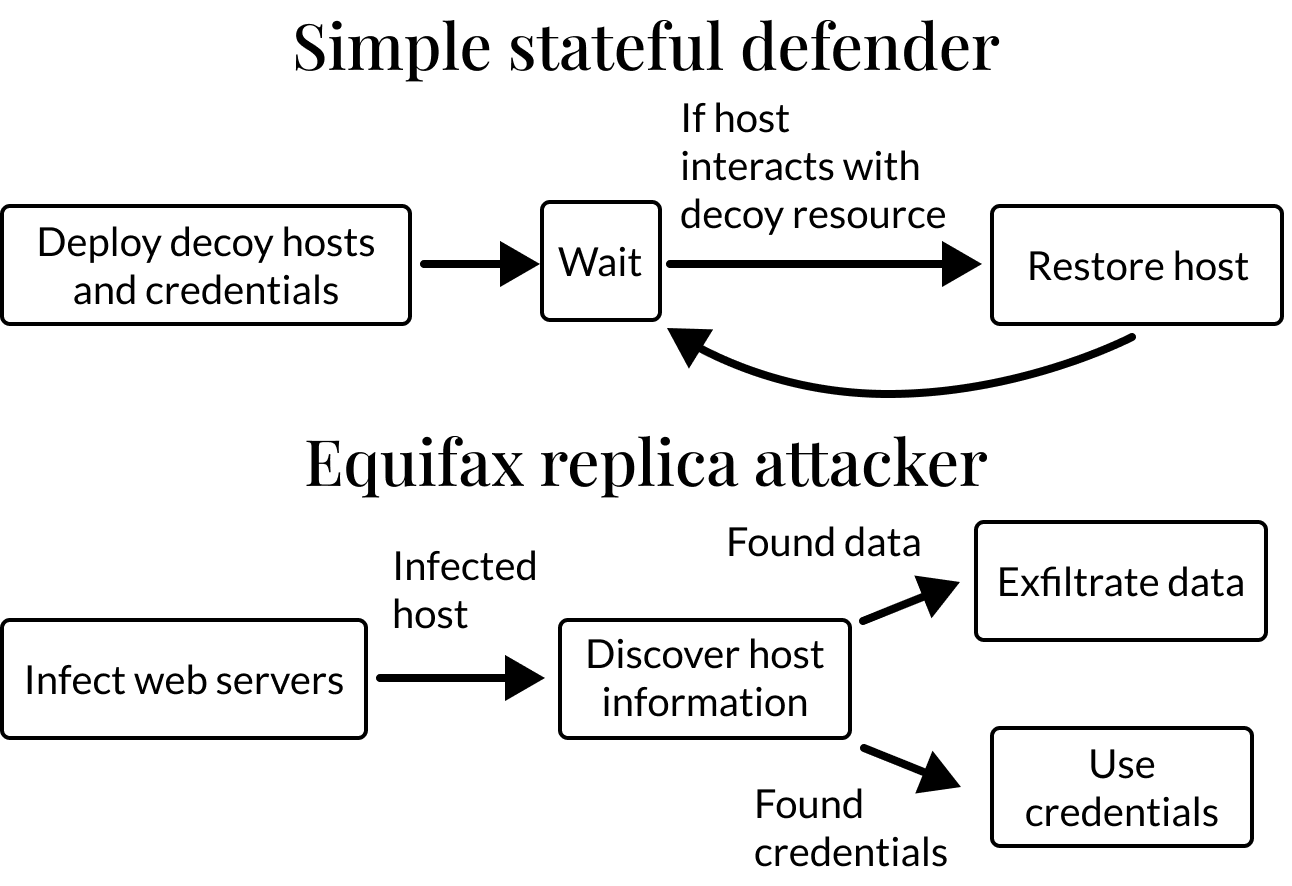}
    \tightcaption{State machines representing the defender and attacker in \secRef{sec:motivation}. The defender deploys decoys, waits for attackers to interact with decoys, and then restores hosts to stop attackers. Similarly, the attacker can be modeled as a state machine to infect web servers, find credentials, and finally exfiltrate data.}
    \label{fig:example_state_machines}
\end{figure}

\subsection{High-level idea}
To put the key ideas of Perry in context, let us revisit from first principles {\em what} the operator was trying to do and {\em why} existing tools failed to allow for the expression of those intents.

Conceptually, the operator has a mental model of  the  attacker and defender logic (\secRef{sec:motivation}) shown in the simple  state machines in \fig{fig:example_state_machines}.
 Here, the defender wants to deploy decoys, wait for attackers to interact with the decoys, and then runs actions to stop the attacker.
Similarly, the attacker runs like a logical  state machine that tries to infect web servers, find credentials, and finally exfiltrate data.

While these intents are easy to describe in this mental model, there is a significant disconnect with existing tools.
Existing frameworks are environment-specific, low-level, and lack any abstraction, making it painful to realize these intents.
For example, the first state of the defender in \fig{fig:example_state_machines} is to deploy decoy hosts and credentials.
But modeling this state in our implementation requires 240 lines of code (163 lines of Ansible code and 77 lines of Python code), which only work in one environment. 
As we discussed earlier, this code is tightly coupled, making it difficult to extend to new environments, capabilities, and strategies. 

To tackle this disconnect, Perry  raises the level of abstraction to express diverse and complex attackers and defenders in an environment-agnostic way. 
In fact, the mental model of the operator described above naturally suggests a convenient  abstraction to capture  attackers and defenders.

Based on this insight, Perry introduces a high-level programming model, explicitly representing attackers and defenders as logical state machines that execute {\em high-level actions} based on {\em high-level observations} of the environment.
In essence, Perry helps the user express high-level intentions of {\em what} they need  to do, decoupling intentions from the {\em how} in low-level tools.  

Next, to make this more concrete, we present a brief sketch of how operators can design deception what-if experiments with Perry's high-level programming model.
Later, in \secRef{sec:detailed_system} and \secRef{sec:illustrative_examples}, we give more concrete  examples.

\subsection{Anatomy of an experiment}
A deception what-if scenario has (1) a deception defense, (2) an attacker, and (3) an environment.

\para{Deception defense} In Perry, deception defense has three stages: A) initialization, set up strategies and deploy deception resources; B) event listeners, strategies subscribe to high-level observations from the telemetry service; and C) event handlers, strategies react to the high-level observations.
For instance, the pseudo code of the stateful deception approach in \figRef{fig:example_state_machines} is (full example in \secRef{sec:illustrative_examples}):
\begin{lstlisting}[language=Python]
# (A) Initialize, deploy deception e.g.,
DeployDecoyHost(important_network)
# (B) Event listeners, detect interactions e.g.,
TelemetryModule.subscribe(
    DecoyHostInteraction,
    my_handler)
# (C) Event handlers, respond to attacker e.g.,
def my_handler(event):
    RestoreHost(event.source_host)
\end{lstlisting}

\para{Attacker} In Perry, attackers are written as explicit state machines.
First, the strategies enumerate each attack phase.
Then, each attack phase is implemented with a mixture of Perry's modules.
For instance, the pseudo code of the Equifax replica attacker in \figRef{fig:example_state_machines} is (full example in \secRef{sec:illustrative_examples}):
\begin{lstlisting}[language=Python]
# (A) Attack phases
match self.attack_stage:
    case InitialAccess:
        self.initial_access()
    case LookAndUseCredentials:
        self.discover_info()
    # ...
# (B) Phase implementation
def initial_access():
    ScanNetwork(external_network)
    # ...
    for path in new_attack_paths
        InfectHost(path)
\end{lstlisting}

\para{Environment} Environments in Perry are specified by a topology and a configuration file.
The topology is specified in Terraform~\cite{terraform}.
The configuration file specifies the host configurations, vulnerabilities, and goals.
For example, the pseudo code of the Equifax environment is:

\begin{lstlisting}[language=Python]
# (A) Host configurations
ConfigureSSHKeys(self.webserver[0],
                 self.databases)
# (B) Vulnerabilities
for webserver in self.webservers:
    SetupVulnerableApacheStruts(webserver)
# (C) Goals
for database in self.databases:
    AddCriticalData(database)
\end{lstlisting}

\section{Detailed system design} \label{sec:detailed_system}
\begin{figure}[tb]
    \centering
    \includegraphics[width=0.45\textwidth]{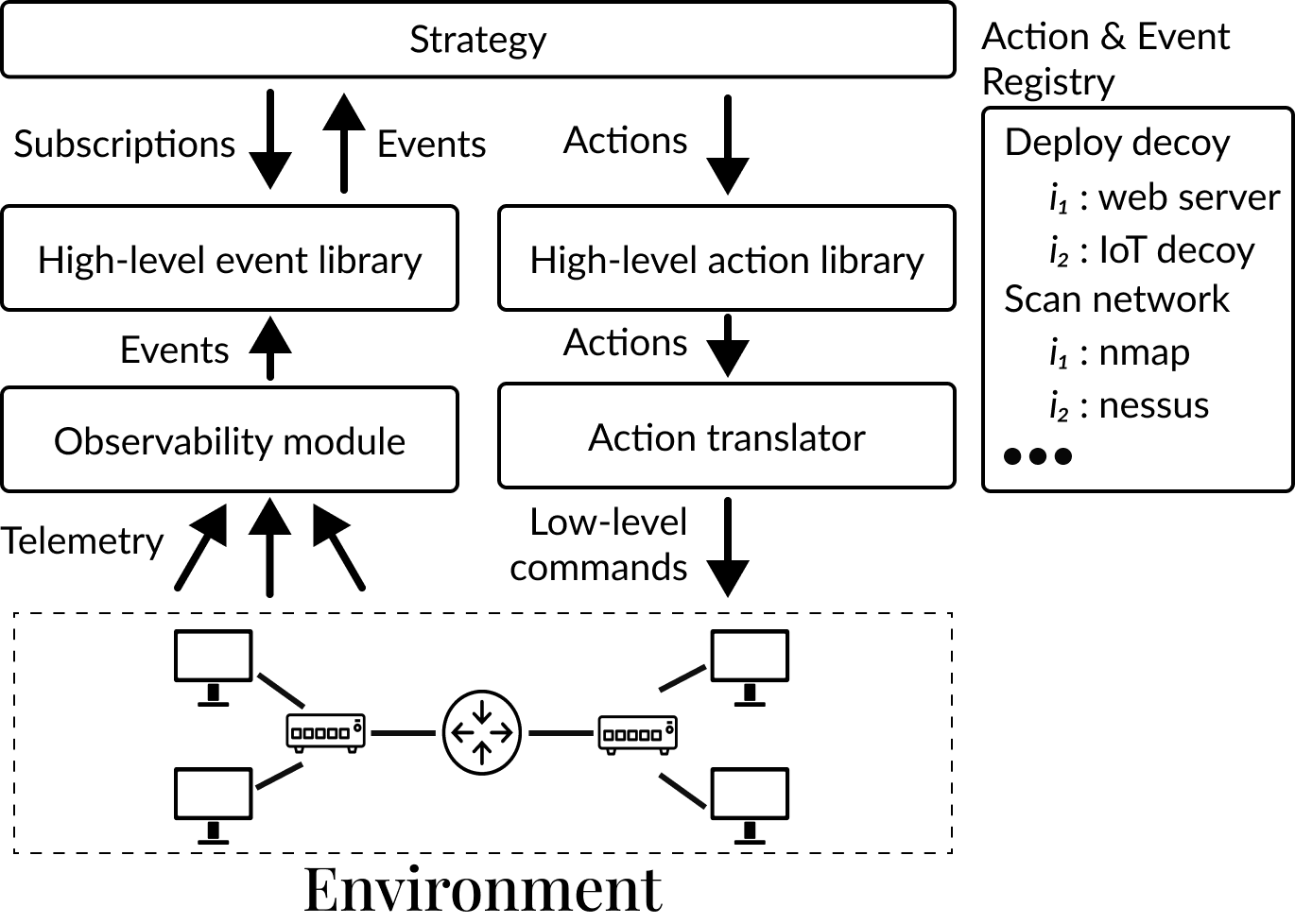}
    \tightcaption{Perry has several modules and services to decouple the high-level intentions of strategies.
    In particular, there is an \actionModule\ to convert high-level actions into low-level actions and a \telemetryModule\ to translate low-level observations into high-level observations.
    The registry contains a list of high-level actions and events with a list of potential implementations.}
    \label{fig:unified_abstraction}
\end{figure}

Next, we describe the key modules' design in Perry: environment state service, \actionModule, \telemetryModule, and attack graph service.
Broadly speaking, strategies subscribe to high-level observations from the \telemetryModule\ and output high-level actions for translation by the \actionModule, shown in \figRef{fig:unified_abstraction}.

\subsection{Environment state service}
\label{sec:detail_env_service}
As shown in \secRef{sec:motivation}, strategies in existing tools are coupled to a single environment.
Rather than strategies having low-level and environment-specific code, in Perry, code that reasons about the environment is delegated to the environment state service.
Perry's modules and strategies query the environment state service for environment information (e.g., check if a credential is fake).
To answer these queries, the environment state service maintains a network information base~\cite{holm2022lore} by subscribing to high-level events from the \telemetryModule\ (\secRef{sec:detail_telemetry_module}).
For example, for attackers, the environment state service subscribes to a new host discovery event and adds hosts to the knowledge base as they are discovered. 

As an illustration, the following snippet shows a simple randomized decoy host and credential deployment using the Perry API. 
The key idea here is that all environment-specific code is delegated to the environment state service:
    
\begin{lstlisting}[language=Python]
public_subnets = EnvService
    .get_subnet(public=True)
# Deploy decoy credentials
for _ in range(0, num_honeycreds):
    deploy_host = EnvService
        .get_random_host(public_subnets)
    decoy = EnvService.get_random_decoy_host()
    DeployDecoyCredential(deploy_host, decoy)
\end{lstlisting}

In the above example, the implementation is now environment-agnostic.
Rather than the code specifying low-level individual IP addresses to deploy decoy credentials, it uses the environment state service to select the high-level intention of the host for deployment of the decoy credentials.

\subsection{\ActionModule}
\label{sec:detail_action_module}
To make it easier to write and modify high-level strategies, we design an \actionModule\ that decouples strategies' high-level actions from their low-level implementations.

Currently, actions in attack and defense emulation tools~\cite{caldera, holm2022lore, enoch2020harmer, atomic_red_team} are specified with low-level commands that have pre and post-conditions.
For example, one command might be an \texttt{nmap} scan script with the post-condition to execute an \texttt{SSH} exploit if a known CVE is found.

The pre and post-conditions have two challenges: they make it difficult to modify high-level strategies, and they complicate the addition of new capabilities because each one requires reconsideration of existing conditions.
For instance, if a user wants to modify an attack strategy to mimic the Darkside~\cite{darkside} APT attack group, the user would have to alter many low-level pre-conditions to match the new high-level specification. 

In Perry, the high-level actions are decoupled from their implementations.
As a result, the \actionModule\ makes it easier to specify, modify, and reason about strategies because users no longer have to reason about low-level code.
Furthermore, the decoupling of implementations makes the \actionModule\ extensible in terms of both adding new high-level actions (e.g., adding a stealthy scan action) and adding new implementations of actions (e.g., a better data exfiltration technique).
For example, in a strategy, we can program an exfiltrate data action as:
\begin{lstlisting}[language=Python]
ExfiltrateData(host_with_data)
\end{lstlisting}

The \lstinline{ExfiltrateData} could have multiple implementations.
In our implementation, we create a list of possible exfiltration paths with the \texttt{SSH} and \texttt{HTTPS} protocols.
Then, we find the shortest exfiltration path.
However, other users can add alternate implementations to include other protocols (e.g., \texttt{DNS}).
\emph{The key takeaway is that users choose the implementation they want to explore without modifying existing strategies.}

\begin{table}[tb]
  \centering
  \caption{Perry’s high‑level actions and their default implementations.}
  \label{tab:action_translation}

  \begin{tabularx}{\linewidth}{@{}>{\raggedright\arraybackslash}l X@{}}
    \toprule
    \multicolumn{2}{c}{\textbf{Defender actions}}\\
    \cmidrule(lr){1-2}
    \textbf{Action} & \textbf{Implementation} \\ \midrule
    Deploy decoy host        & Creates a decoy host with options such as a vulnerable service, fake data, or fake users.\\
    Deploy decoy credential   & Adds fake SSH credentials to a server and optionally links them to a decoy host.\\
    Deploy decoy data         & Generates files containing fake sensitive data.\\
    Deploy decoy users        & Creates a decoy user with a weak password.\\
    Deploy honey service      & Configures an interactive fake network service.\\
    Restore host              & Restores the host to a clean snapshot.\\
    Shutdown host             & Shuts down the host.\\
    \addlinespace
    \multicolumn{2}{c}{\textbf{Attacker actions}}\\
    \cmidrule(lr){1-2}
    \textbf{Action} & \textbf{Implementation} \\ \midrule
    Discover local information & Searches directories for files and credentials.\\
    Scan                       & Uses \texttt{nmap} to find vulnerable network services.\\
    Lateral move               & Searches an exploit database and executes suitable exploits to pivot to another host.\\
    Escalate privilege         & Searches an exploit database and executes suitable exploits to gain higher privileges.\\
    Exfiltrate data            & Finds the shortest path to the attacker’s host and exfiltrates the data.\\
    \bottomrule
  \end{tabularx}
\end{table}

We also design a library of common high-level deception and attack actions that are portable across environments in \tableRef{tab:action_translation}.
For instance, the data exfiltration action (1) uses the attack graph service to identify the shortest exfiltration path and (2) uses the environment state service to choose the correct protocol (e.g., \texttt{HTTP}) for data exfiltration.

\subsection{\TelemetryModule}
\label{sec:detail_telemetry_module}
Similar to actions, defenders and attackers need to reason about the state of an environment (e.g., host used a decoy credential, infected a new host, found critical data).
As seen in \secRef{sec:motivation}, existing tools expose low-level events (e.g., system calls, network tracing, outputs of a bash command, etc.) and require environment-specific information (e.g., IP address of a decoy host).
To make it easier to write and modify high-level strategies, we design an environment-agnostic \telemetryModule\ inspired by Zeek~\cite{paxson1999bro} that decouples strategies' high-level observations from their low-level events.

Current attack and defense emulation tools require implementations to have parsers for low-level telemetry data (e.g., network traces, command outputs)~\cite{caldera, holm2022lore, enoch2020harmer, elasticsearch, ibmSecurityQRadar}.
As a result, strategy implementations become (1) hard to modify because they have large amounts of low-level code to process the data and (2) hard to extend because they are tightly coupled to the type of telemetry data.

Instead of strategies containing complex logic to interpret low-level telemetry, Perry enables strategies to subscribe to high-level observations.
The \telemetryModule\ translates low-level observations into these high-level observations.
For example, in Perry, a strategy can subscribe to a high-level observation that a host is using a decoy credential:
\begin{lstlisting}[language=Python]
TelemetryModule.subscribe(
    HostUsedDecoyCredential, handle_decoy_host_interaction)
\end{lstlisting}

\begin{table}[tb]
  \centering
  \caption{Perry’s high‑level events and their default implementations.}
  \label{tab:event_translation}

  % l = ragged‑right first column, X = stretchable second column
  \begin{tabularx}{\linewidth}{@{}>{\raggedright\arraybackslash}l X@{}}
    \toprule
    \multicolumn{2}{c}{\textbf{Defender events}}\\
    \cmidrule(lr){1-2}
    \textbf{Event} & \textbf{Implementation} \\ \midrule
    Decoy host interaction   & Network‑trace rules that trigger when non‑decoy hosts establish connections to decoy hosts.\\
    Credential interaction   & eBPF rules detect decoy credential use.\\
    SSH connection           & Network‑trace rules that trigger when an SSH session is established.\\
    \addlinespace
    \multicolumn{2}{c}{\textbf{Attacker events}}\\
    \cmidrule(lr){1-2}
    \textbf{Event} & \textbf{Implementation} \\ \midrule
    Found host               & Parses scan outputs to find new hosts.\\
    Found network services   & Parses scan outputs to find new services.\\
    Found credential         & Parses command outputs to find credentials.\\
    Data found               & Searches file contents for sensitive keywords.\\
    Exfiltrated data         & Triggers when data is successfully exfiltrated.\\
    Infected new host        & Triggers when malware is correctly installed.\\
    Got root access          & Triggers when malware is installed as root user.\\
    \bottomrule
  \end{tabularx}
\end{table}

Here, we decouple the high-level intention of detecting hosts using decoy credentials from the low-level implementation.
Consequently, we can quickly explore new telemetry techniques without modifying the strategy.
For instance, if we implement a new anomaly detection algorithm to output \lstinline{HostUsedDecoyCredential} events, it works with the existing strategy.
We design a library of common high-level deception and attacker events, shown in \tableRef{tab:event_translation}.

\subsection{Attack graph service}
\label{sec:detail_attack_graph_service}
Attack and defense strategies often have to reason about vulnerabilities in networks.
For example, an attack may want to target a specific host or get privileged access to an already infected host.
Defenders too may want to proactively explore possible future attack steps.  
As seen in \secRef{sec:motivation}, existing tools only offer low-level methods for reasoning about potential targets in a network.
To make it easier to reason about vulnerabilities, we design an attack graph service.

To reduce the effort required to express  attackers with sophisticated strategies, we revisit a classic but effective idea---attack graphs~\cite{ou2005mulval}.  
For example, if a strategy wants to infect a specific host on a network, it can query for potential infection methods from a service:
\begin{lstlisting}[language=Python]
paths = AGService.get_paths_to_host(
    attacker_host, target_host)
for path in paths:
    events = LateralMove(path)
\end{lstlisting}

Because attackers can only have partial knowledge about the environment, the attack graph service continually updates itself as new information is discovered using the environment state service.
For example, if credentials were discovered on a host, the attack graph service will identify new paths that infect hosts.
These paths can directly be executed by the \actionModule.
As a result, the attack graph service removes complex logic from the strategy and creates a modular query so that other strategies can also request infection methods. 

\subsection{Attacker service}
Our attacker service enables us to model realistic attackers. 
Attackers in Perry use C\&C servers, install malware agents on hosts, scan networks, and execute real exploits.
The high-level attacker actions in \tableRef{tab:action_translation} use the action translation module to translate these high-level actions into common low-level attack tools.
For instance, the lateral movement high-level action either uses credentials or exploits to install malware agents on the target host.
The exploits are currently selected from a small built-in library; however, Perry also supports an optional module that can use Metasploit's larger exploit library.
In addition, we envision the community creating custom implementations of these actions, such as a lateral movement implementation that uses LLMs to generate the exploits~\cite{deng2023pentestgpt, singer2025feasibility}.  

% One benefit of Perry is that even with these high-level abstractions 

% The attack graph service is 

% Not human limitation -> future may be autonomous

\section{Illustrative example} \label{sec:illustrative_examples}
In this section, we show how using Perry can simplify the implementation for the stateful deception approach and Equifax attacker in \fig{fig:example_state_machines}.
These examples illustrate the expressiveness of how deception strategies, attackers, and environments are specified in Perry.
Then, we illustrate how operators can extend Perry through qualitative examples.

Recall that the simple stateful strategy in \figRef{fig:example_state_machines}.
The simple stateful strategy first deploys deceptive resources, then restores any hosts that react with them.
In Perry, we express this as:
\begin{lstlisting}[language=Python]
# (A) Initialize, deploy deception
for _ in range(0, num_decoys):
    subnet_to_deploy = EnvService
        .get_random_subnet()
    DeployDecoyHost(subnet_to_deploy)
for _ in range(0, num_honeycreds):
    deploy_host = EnvService.get_random_host()
    decoy = EnvService.get_random_decoy_host()
    DeployDecoyCredential(deploy_host, decoy)
# (B) Event listeners, listen for interactions
TelemetryModule.subscribe(
    HostInteractedWithDecoyHost, handle_decoy_host_interaction)
# (C) Event handlers, react to attacker
def handle_decoy_host_interaction(event):
    RestoreServer(event.source_host)
    RestoreServer(event.target_host)
\end{lstlisting}

In the initialization phase, we use the environment state service to select random subnets to deploy decoy hosts to and select random hosts to deploy decoy credentials to. 
The \actionModule\ then translates the  deploy decoy and credential actions into their low-level commands.
We use the \telemetryModule\ to subscribe to a key event, a decoy host interaction.
Next, we specify the event handlers to react to these interactions by restoring both the decoy host and the host that interacted with the decoy host.

We can similarly use the Perry abstractions to express the  attacker from \fig{fig:example_state_machines}:
\begin{lstlisting}[language=Python]
# (A) Attack phases
match self.attack_stage:
    case InitialAccess:
        self.initial_access()
    case LookAndUseCredentials:
        self.discover_info()
    case ExfiltrateData():
        self.exfiltrate_data()
# (B) Phase implementation
def initial_access(self):
    ScanNetwork(EnvService.get_subnets())
    # Try to infect all hosts on external network
    hosts = EnvService.get_hosts()
    attack_paths = AGService.get_all_paths()
    for path in attack_paths:
        LateralMove(path)
    self.state = LookAndUseCredentials
# ...
\end{lstlisting}

We enumerate each of the attack phases in the Equifax replica strategy: 1) gain initial access to the network, 2) find and use any credentials, and 3) exfiltrate any data.

Then, we implement each phase, starting with initial access.
The initial access phase first uses the environment state service and \actionModule\ to scan the external subnets.
Next, we use the attack graph service and \actionModule\ to infect any vulnerable servers. We create the exfiltrate data phase by first using the environment state service and \actionModule\ to find critical data.
If any critical data is found, the \telemetryModule\ will update the environment state service.
As a result, after we execute find critical data, we check if any data is on the host and exfiltrate it if so.

Overall, we are able to reduce the total lines of code from 467 to 44 and from 1596 to 94 for the defender and attacker respectively (\secRef{sec:eval_rapid_exploration}).
Even more importantly, Perry gives an intuitive high-level abstraction for the user.
In the next section, we discuss the detailed design of the Perry modules.

\para{Environment expressiveness}
In Perry, environments are also specified at a high level.
The network topologies are defined with Terraform, an industry standard for specifying virtual network topologies~\cite{terraform}.
In addition to the topology, environments have a Python specification with the environments vulnerabilities, host configurations, and key assets.
Below is a snippet of the specification of the Equifax-inspired environment used in \secRef{sec:eval}:

\begin{lstlisting}[language=Python]
# (A) Setup ApacheStruts Vulnerability
for web_server in self.webservers:
    SetupStrutsVulnerability(web_server, port=443)
# (B) Setup credentials
cred_web_server = self.webservers[0]
for db in self.database_hosts:
    SetupServerSSHKeys(
        cred_web_server.ip, 
        cred_web_server.users[0], 
        db.ip, 
        db.users[0]
    )
# (C) Setup database data
for db in self.database_hosts:
    AddData(db, database.users[0])
\end{lstlisting}
In this example, first the vulnerable ApacheStruts web service is installed on the two web servers.
Then, one web sever is given SSH credentials to all of the databases.
Finally, each data is added to each database for the defender to protect. 

\begin{table}[tb]
  \centering
  \caption{Attack strategies implemented with Perry.  Equifax and Darkside replicas follow public reports\cite{equifax_report,darkside}. 
  DFS Movement, Targeted, and Persistent strategies have different priorities. 
  See Appendix~\ref{sec:appendix_attackers} for details.}
  \label{tab:eval_attack_strategies}

  \begin{tabularx}{\linewidth}{@{}>{\raggedright\arraybackslash}l X@{}}
    \toprule
    \textbf{Attack strategy} & \textbf{Description} \\ \midrule
    DFS Movement      & Laterally moves through the network using a depth‑first‑search (DFS) traversal of the attack graph.\\
    Equifax replica   & Mimics the 2017 Equifax breach: gains initial access, harvests credentials, and exfiltrates data.\\
    Targeted          & Starts with prior knowledge of the network and uses it to prioritise high‑value hosts in the attack graph.\\
    Persistent        & Infects vulnerable hosts and keeps them as back‑door footholds for long‑term access.\\
    Darkside replica  & Emulates the DarkSide APT: gains initial access, spreads widely, then executes its final objective.\\
    \bottomrule
  \end{tabularx}
\end{table}

\section{Implementation}
\label{sec:implementation}
In this section, we describe the implementation of Perry. 
First, we discuss how Perry instantiates environments using OpenStack~\cite{openstack}. % an open source cloud infrastructure.
Then, we discuss Perry's custom Python framework for defender programming. 
Last, we describe Perry's attacker implementation.

\parabf{Environment instantiation}
Perry uses an environment specification to instantiate an environment in an OpenStack cloud.
We use OpenStack because it can run on commodity hardware, is open-source, and scalable~\cite{openstack}.
In addition, we implement components to reduce the time for environments to be reinstantiated by caching environment information.

In Perry, we define an \textit{environment} as a tuple $\langle \mathit{Topology}, \mathit{Assets}, \textit{Host configuration}, \textit{Vulnerabilities}\rangle$. 
The \textit{topology} is the routers, switches, firewalls, and hosts of the network. 
The \textit{assets} are objects such as critical data. 
The \textit{host configuration} are the configurations of hosts such as deploying a database or creating a user. 
The \textit{vulnerabilities} misconfigure hosts or setup vulnerable software versions.

The environment's network topology is specified using Terraform Language~\cite{terraform}, an ``infrastructure as code'' language.
The assets, host configurations, and vulnerabilities, are specified using an internal Python SDK.
The internal Python SDK is designed to be extensible for users to add additional assets, configurations, and vulnerabilities.

Environments are instantiated in two phases: setup and launch.
The setup phase deploys and configures the network and hosts in OpenStack.
Next, assets, host configurations, and vulnerabilities are configured through a mixture of Python and Ansible~\cite{ansible}.
Once the environment is configured, Perry snapshots each host to save time launching the environment.
During the launch phase, Perry restores the network topology to its original state.
Then, Perry instantiates each host on the network using the saved snapshots.
The VM snapshots are RAW images which are portable to most cloud providers (e.g., AWS, Google Cloud).\footnote{
As future work, we plan to support several major cloud providers in the open-source repository.
}

\parabf{Defenders}
We implement  Perry's abstractions for  defenders' deception approaches with   a custom Python framework.
Each defender is specified with a configuration that contains the strategy, capabilities, and \telemetryModule.
The capabilities specify the actions available to the defender and the budget for each of these actions (e.g., a budget of 10 decoy hosts).

The defender's \actionModule\ uses orchestrators to enable Perry to work in multiple clouds such as OpenStack~\cite{openstack}, or AWS~\cite{aws}.
In this paper, we implement an OpenStack orchestrator.
For a high-level action, the orchestrator will use the OpenStack SDK to execute the low-level actions.
We use Ansible to execute the remaining low-level actions that are not specific to the cloud environment. 
We implement the defender's \telemetryModule\ with Python and Elasticsearch~\cite{elasticsearch}.
The \telemetryModule\ continually queries the database and has rules that raise high-level observations.

\parabf{Attackers}
The attacker service uses Caldera, an open-source C\&C server to communicate and send commands to infected hosts.\footnote{
Other C\&C server's such as Mythic~\cite{merlin} or CobaltStrike~\cite{cobalt} could also be used.
}
The \actionModule\ translates tasks from Perry's high-level attack strategies into low-level commands (e.g., Shell code, Python scripts) that are then executed by Caldera~\cite{caldera}.
We implement the \telemetryModule\ by considering the outputs of each low-level action as low-level events.
For each low-level action, we create rules that parse these low-level outputs and raise high-level observations.
Similar to the defender, we implement a library of common high-level attack actions and events shown in \tableRef{tab:action_translation} and \tableRef{tab:event_translation}.

%%%% Section Macros %%%%
\newcommand{\LOCOfInfectHost}[0] {139}
\newcommand{\LOCOfParsingActions}[0] {88}
\newcommand{\LOCOfKnowledgeBaseObjects}[0] {178}
%%%%%%%%%%%%%%%%%%%%%%%%

\section{Evaluation} \label{sec:eval}
In this section, we first evaluate how Perry's abstractions reduce the implementation effort for a wide variety of deceptive defenders and attackers. Second, we discuss how Perry is extensible to enable operators to  specify  variations of existing  strategies and explore new defense and attack capabilities as they become available.
Third,  we show how Perry can shed light on the efficacy of various deception defenses and their cost-benefit tradeoffs. 

\begin{table}[tb]
  \centering
  \caption{Deception strategies implemented with Perry. See Appendix~\ref{sec:appendix_defenders} for details.}
  \label{tab:eval_defense_strategies}

  \begin{tabularx}{\linewidth}{@{}>{\raggedright\arraybackslash}l X@{}}
    \toprule
    \textbf{Deception strategy} & \textbf{Description} \\ \midrule
    Basic honeypot      & Randomly deploys honeypots across the network.\\
    Mixed deception     & Randomly deploys honeypots and fake credentials across the network.\\
    Layered deception   & Deploys honeypots, decoy credentials, and fake data on each honeypot; the decoy credentials are valid for those honeypots.\\
    Simple stateful     & Deploys honeypots with basic telemetry that alert on host interaction.\\
    \bottomrule
  \end{tabularx}
\end{table}

To execute these analyses we select representative and realistic attackers, deception strategies, and environments.
For attackers, we design and implement five attack strategies shown in \tableRef{tab:eval_attack_strategies}.
The attackers replicate real-world attack tactics and APTs from public reports~\cite{equifax_report, darkside, opmBreach, vulnerablePrinters}.
We picked four deception strategies shown in \tableRef{tab:eval_defense_strategies} (and implement additional deception strategies in the next sections) inspired by common deception strategies~\cite{moore2016malwareHoneypot, pa2016iotpot, amnesia_honey_db, real_world_honeypot}.

\subsection{Reduction in  implementation effort}
\label{sec:eval_rapid_exploration}
We start with measuring the reduction in implementation effort in terms of lines of code (LOC) as a measure of specification and debugging complexity.
Then we highlight the benefit of Perry in avoiding the environment-specific effort that is a key weakness of previous tools.
We also show how the Perry API can enable more robust LLM-based code generation of strategy specifications.

% Environments table
\begin{table}[tb]
\caption{We implement five environments in Perry. The Equifax-inspired and Colonial pipeline-inspired are based on real attacks~\cite{equifax_report, colonial_pipeline_techtarget}. The Chain and Star environments are in prior deception papers~\cite{mirage, ringNetwork2013technique, ferguson2021_deception_psychology}.}
    \label{tab:eval_environments}
    \centering
    \begin{tabular}{>{\raggedright\arraybackslash}p{0.22\linewidth}>{\raggedright\arraybackslash}p{0.7\linewidth}}
    \toprule
         \textbf{Environment}& \textbf{Description}\\
         \midrule
         Equifax-inspired \newline(50 hosts)& A replica of Equifax network (same topology, services, and vulnerabilities)~\cite{equifax_report}. The goal is to exfiltrate all 48 databases.\\
         \hline
         Colonial Pipeline-inspired \newline(45 hosts)& An environment inspired by the Colonial Pipeline breach~\cite{colonial_pipeline_techtarget} and other ICS attacks~\cite{lee2017crashoverride, sheddingLight}. The goal is to gain access to 15 physical actuators.\\
         \hline
         Chain \newline(25 hosts)& Each host has credentials to another host~\cite{mirage, ringNetwork2013technique}.
The goal is to exfiltrate critical data on each host.\\
         Enterprise \newline(20 hosts)& A tree topology, sometimes used in enterprise networks~\cite{ibmTreeNetwork, ciscoEnterpriseNetwork}. There is one external network and two networks for each floor of a building.\\
         \hline
         % Dumbbell \newline(30 hosts)& An environment with 2 networks, one network with 15 webservers and another with 15 databases~\cite{dumbbellNetwork}. The goal is to exfiltrate all data in the network.\\
         % \hline
    Star \newline(25 hosts)&A single network where all hosts have vulnerabilities and contain critical data~\cite{ferguson2021_deception_psychology}.\\
    \bottomrule
    \end{tabular}
\end{table}

\begin{figure}[tb]
    \centering
    \includegraphics[width=0.42\textwidth]{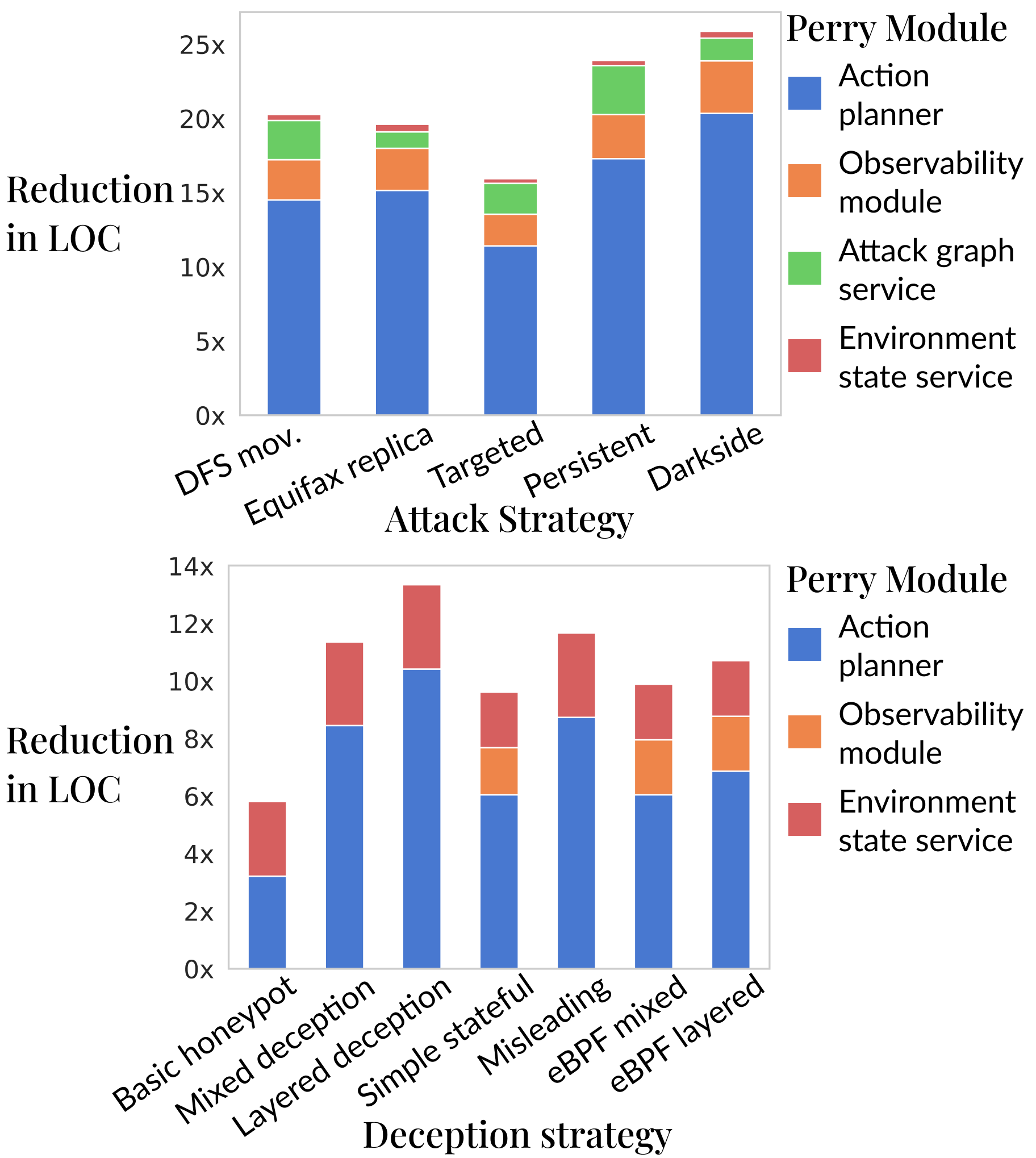}
    \tightcaption{Across all defender strategies, Perry's abstractions reduce the LOC by 6.0-12.7$\times$. Across all attacker strategies, Perry's abstractions reduce the LOC by 13.1-22.2$\times$. Most of the reduction is from the \actionModule.}
    \label{fig:eval_loc}
    \vspace{-0.2cm}
\end{figure}

% Environment agnostic attackers and defenders
% LOC reduction
\para{LOC measurement} 
 While we acknowledge lines-of-code (LOC) is not a perfect  measure of implementation effort, it is still a useful proxy  metric for effort, code readability,  and debugging complexity. To this end,  
we count and compare the LOC of implementing attackers and deception approaches with and without using Perry's abstractions (\figRef{fig:eval_loc} and Appendix \ref{sec:appendix_loc}).
We also  break down how much each Perry module contributed to the reduction. 

For attack implementations, we compare Perry against implementations using  low-level Caldera actions. 
In \figRef{fig:eval_loc}, we show Perry reduces LOC of the attacker implementations 16.9--26.9$\times$.
The \actionModule\ provided the biggest benefit for LOC reduction  by offloading the complex logic to execute low-level calls to the Caldera SDK.
The Darkside strategy had the largest savings in LOC, with a 26.9$\times$ reduction.

For defense strategies, we compare Perry vs.\   implementations  expressed using a combination of Python code and Elasticsearch.
In \figRef{fig:eval_loc}, we show Perry reduces LOC of the deception  implementations by 6.8--14.3$\times$.
Again, the \actionModule\ provides the largest benefit 
 by avoiding  the complex interactions with the OpenStack and Ansible SDKs, shown in \figRef{fig:eval_loc}.
The second highest benefit stems from  the environment state service,
 as  deception strategies often need to reason about the current state of the  environment.

\begin{figure}[tb]
    \centering
    \includegraphics[width=0.35\textwidth]{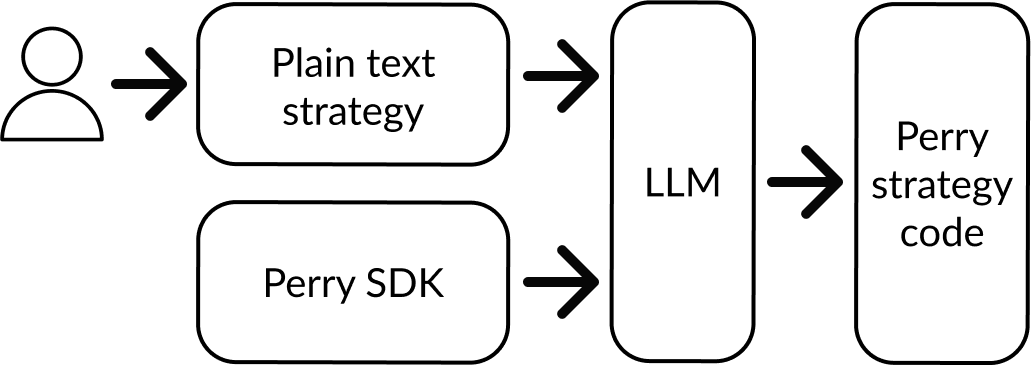}
    \tightcaption{First an operator specifies a plain text description of the deception strategy. Then, both the strategy description and the Perry SDK is given to the LLM.}
    \label{fig:eval_llm_translation}
    \vspace{-0.2cm}
\end{figure}

\para{Validating portability}
We consider  five environments in \tableRef{tab:eval_environments} based on real attacks~\cite{equifax_report, colonial_pipeline_techtarget} and environments in prior deception studies~\cite{mirage, ringNetwork2013technique, ferguson2021_deception_psychology}.
In previous frameworks, users would need to manually reimplement each attack and defense strategy for each environment. That is, effectively  rewriting each attack strategy five times and each defense strategy five times because the implementations are coupled to environment specifics.
In contrast,  using Perry the strategies are environment agnostic and we  only  need to implement  each attack and deception strategy once. We validated (\secRef{sec:eval_findings})  that all the strategies in \tableRef{tab:eval_attack_strategies} and \tableRef{tab:eval_defense_strategies} work seamlessly in all five environments in \tableRef{tab:eval_environments}.

\para{LLM-based code generation}
Given recent advances in LLM-based  code generation, a  natural question is, if and how, the aforementioned effort can be further reduced using LLMs.  
As a first-step, we evaluate LLMs ability at generating three deception defenses.
To this end, we consider two settings: (1) prompting an LLM to produce code  using  current tools, low-level APIs, commands, and (2) providing the LLM with Perry APIs and then prompting it to produce code using this API (\figRef{fig:eval_llm_translation}).
For both settings, we use  Sonnet 3.7 Thinking as the LLM. 
The prompts we use are in \appRef{app:llm_prompts}.

We execute the code to detect errors and also manually review it  to validate the implementation.   
For all three deception strategies, LLMs are able to correctly translate the strategy with Perry's abstractions.
However, the code produced in low-level APIs is incorrect and do not execute without errors. 
We validated the Sonnet 3.7-generated Perry code against the DFS attacker in the Equifax Large environment and find that the behavior is comparable to the manually written defense code.

\subsection{Evaluating Perry's extensibility}
\label{sec:eval_perry_extensions}

With respect extensibility.  operators may want to consider: (1) {\em variations} of the previously implemented strategies and (2) 
{\em new capabilities} for both attacks and defenses. We illustrate how Perry simplifies such extensions.

\para{Exploring variations of strategies}
\label{sec:eval_perry_mutations}
As an illustrative case study, we consider a scenario where  the operator wants to evaluate two extensions: (1)  intentionally leaking wrong  topology information to the adversary  and (2) 
 allocating the deception budget differently  in a more critical section of the network.   
Implementing these extensions to the layered approach (i.e. misinformation and alternative placement) required adding less than 10 lines of code and deleting 3 lines.
In contrast, doing so with the non-Perry scripts would be significantly harder to express and debug due to  environment-specific modifications.

\para{Defender capabilities}
As an illustrative case study, we consider the case where an operator wants to explore the benefits of fine-grained host telemetry tools~\cite{eBPF, sysflow} to  identify when attackers interact with decoy resources.  
Specifically, we consider SysFlow~\cite{sysflow} to collect system calls using eBPF~\cite{eBPF}. This required us to 1) create an Ansible playbook to install SysFlow on each host (26 LOC); 2) add  a Python module to configure  SysFlow  to send the data to the Elasticsearch database (28 LOC); 3) modify  environment files to install SysFlow on all hosts (6 LOC); and 4) add an observation module that uses the system calls to detect when attackers use decoy credentials or interact with decoy hosts (42 LOC).

Note that this implementation effort is a one-time cost to the \telemetryModule\ and is independent of the environments we want to consider. 
In this case, the new telemetry provides the same high-level events so no new code is needed for the deception implementations. 
After implementing the new SysFlow-based observation module, it can interchangeably be swapped with the original observation module in \secRef{sec:eval_rapid_exploration} with a single change to the  configuration file.\footnote{If the new telemetry generates new high level events, then the deception code should be correspondingly updated with new event handlers.}

\para{New attacker capabilities}  Attackers expressed using the Perry APIs can be extended similar to the defender. 
Furthermore, Perry can also to support attacker code that does not use the Perry API. 
For instance, semi- or autonomous LLM-based attackers are starting to show early promise in CTF style challenges~\cite{cyberseceval3}.
It is easy to plug in such emerging attacker capabilities  as well.
The effort to support a new autonomous attacker in Perry was quite minimal; we just need to create a new attack strategy that prompts the LLM for bash commands to execute and respond with the results (52 LOC). 

\subsection{What-If scenarios and interesting findings}
\label{sec:eval_findings}
Now, we show how  Perry can help operators shed light on  the efficacy of different deception approaches, uncover subtle tradeoffs, and identify counterintuitive findings.

\para{Environments and setup} Perry's emulation layer is flexible  and can implement a wide range of  environments including those considered in prior work~\cite{mirage, ringNetwork2013technique, ferguson2021_deception_psychology, ciscoEnterpriseNetwork, dumbbellNetwork} described in Table \ref{tab:eval_environments}. 
In Appendix \ref{sec:app_micro_benchmark}, we benchmark the time it takes to setup and launch the different  environments. At a high level, we find that the one-time setup cost is manageable for a new environment and the time to launch is also reasonable to enable the scale of experiments needed. 

Once we have this setup,  we can systematically evaluate multiple attack strategies (Table \ref{tab:eval_attack_strategies}) against multiple  deception strategies (Table \ref{tab:eval_defense_strategies}) in several environments.  

\begin{figure}[tb]
    \centering
    \includegraphics[width=0.48\textwidth]{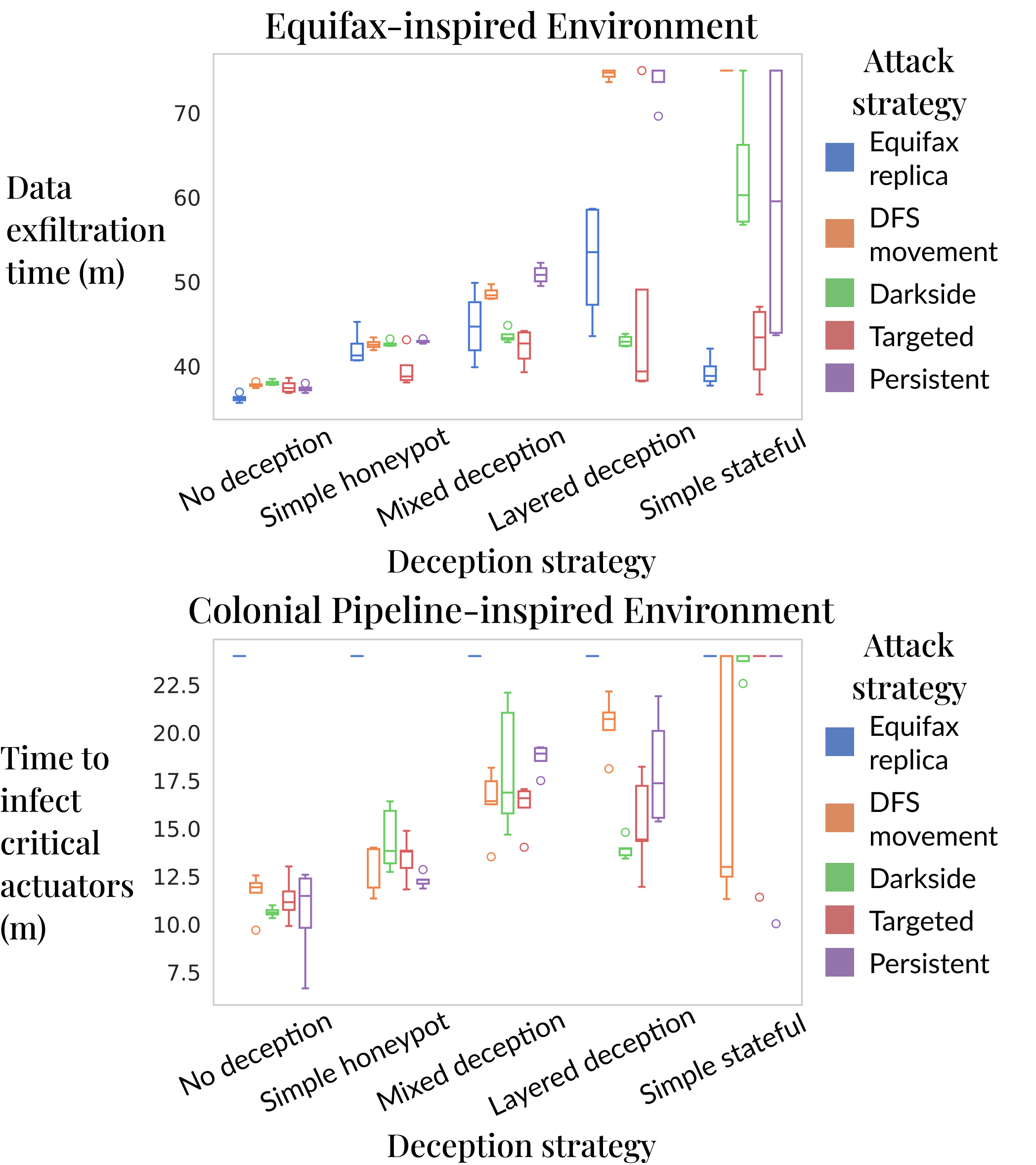}
    \tightcaption{Time metrics for how long attackers took to achieve their goal against each deception strategy.
    The efficacy of deception strategies vary significantly across attackers and environments.}
    \label{fig:eval_time_summary}
    \vspace{-0.2cm}
\end{figure}

\para{Illustrative findings} In what follows, we highlight some interesting findings. 
We do not intend these as conclusive statements about the state of deception or universal facts  across all future  environments. 
We use these as illustrative examples of the types of what-if analysis  tradeoffs operators can uncover.
For each result, we run the experiment 5 times and report the averages across the runs. 
In the interest of brevity, for some  results we only focus on the environments inspired by the Equifax and Colonial Pipeline incidents. 

\begin{figure}[tb]
    \centering
    \includegraphics[width=0.35\textwidth]{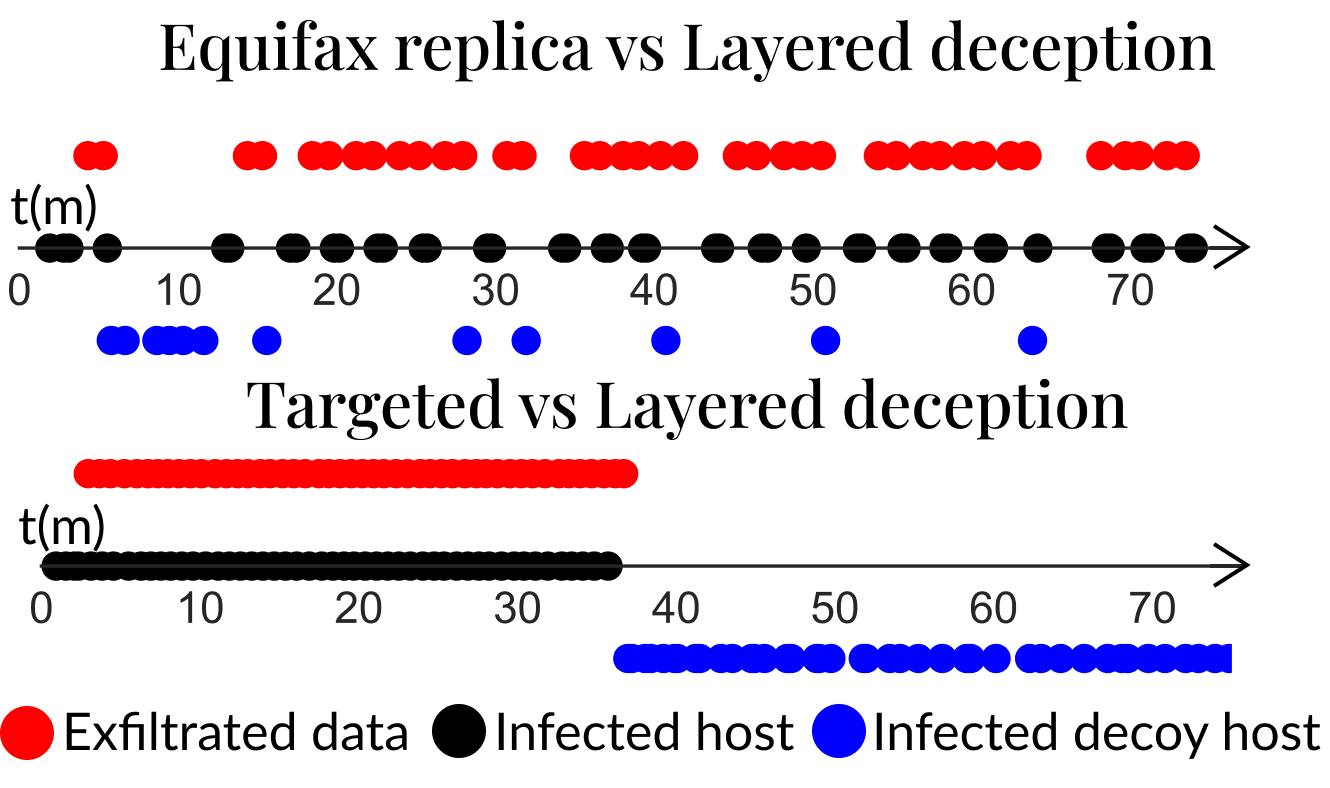}
    \caption{Two example timelines in the Equifax-inspired environment: the Equifax replica attacker and Targeted attacker against the layered deception defense.
    The Targeted attacker uses the prior knowledge of the network to avoid the deception defense.}
    \label{fig:equifax_timeline}
    \vspace{-0.2cm}
\end{figure}

\begin{figure}[tb]
    \centering
    \includegraphics[width=0.35\textwidth]{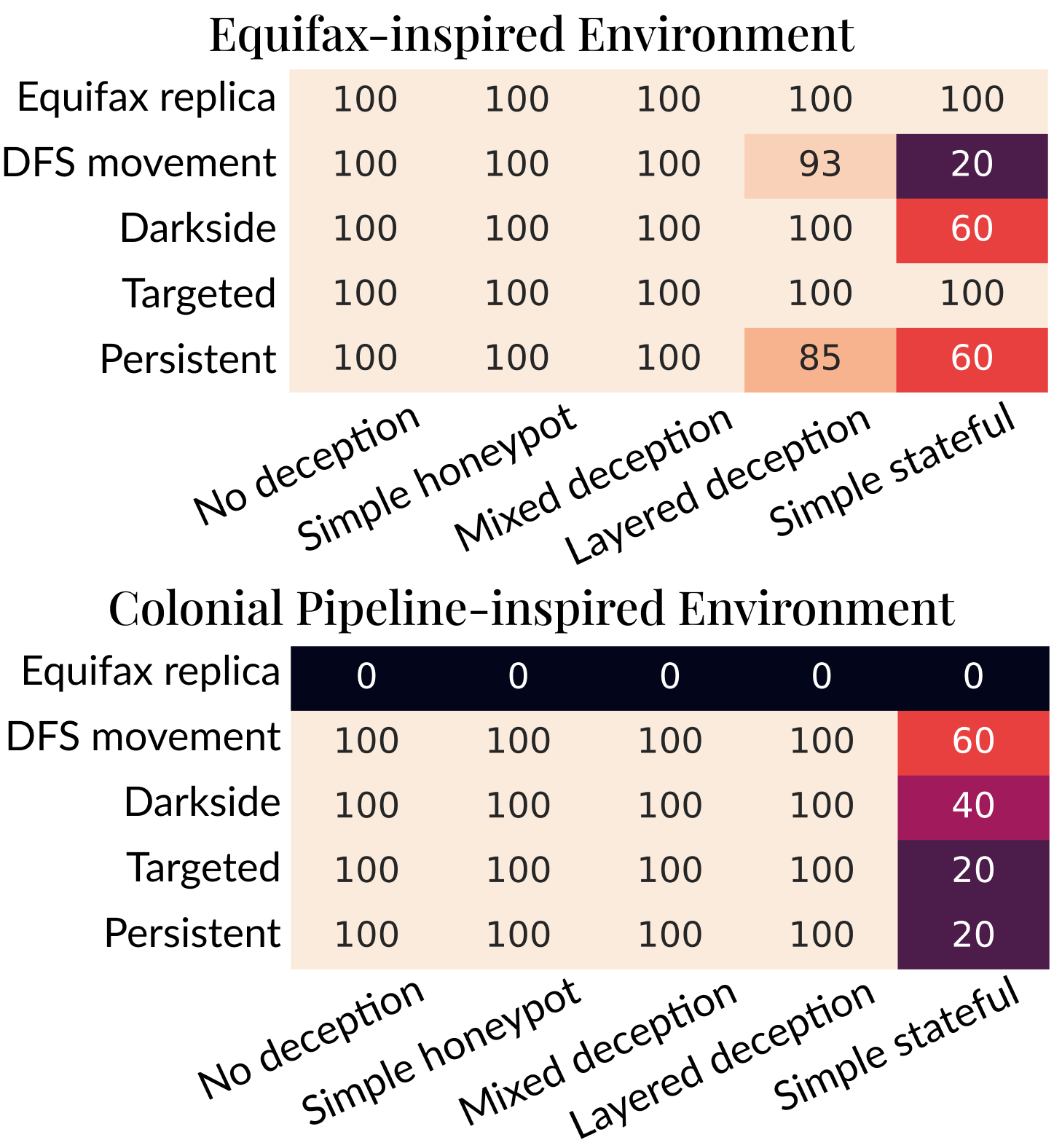}
    \caption{Mean of percent files exfiltrated in the Equifax environment and mean of critical actuators infected in the \cpEnvShort. 
    Only stateful deception strategies that react to the system can thwart attackers.}
    \label{fig:eval_goal_heatmap}
    \vspace{-0.2cm}
\end{figure}

\smallskip \noindent {\bf F1: Value of deception depends on environment  details} \\
In the \equifaxEnvShort\, we consider: (1) binary success metric if  the attacker was able to exfiltrate all data and   (b) the time  to exfiltrate all data. 
In the \cpEnvShort, similarly we use a  binary  metric  if   the attacker was able to infect all critical actuators (\figRef{fig:eval_goal_heatmap}) and the time the attacker took to infect all actuators (\figRef{fig:eval_time_summary}).\footnote{For both time metrics, if the attacker failed, we mark the time as the maximum for the environment.}

We find that deception emulation experiments have many complex and dynamic behaviors that highlight the importance of tools like Perry to directly measure them.
For instance, in the \equifaxEnvShort, a layered deception strategy delays the Equifax replica attacker by 49\%, but delays the targeted attacker by an average of 2.5\% (\figRef{fig:eval_time_summary}).
We illustrate how a Targeted attacker avoids deception with two example timelines in \figRef{fig:equifax_timeline}.
The Targeted attacker uses prior knowledge of the network topology and only interacts with deceptive resources after exfiltrating all of the data.

Furthermore, efficacy of deception strategies vary across environments, emphasizing the need for operators to evaluate deception for their use case.
For example, in the \cpEnvShort\ a mixed deception strategy delays an attacker on average 18\% more than a layered strategy (\figRef{fig:eval_time_summary}), but in the \equifaxEnvShort, both the mixed deception  and layered strategy have the same efficacy.

\smallskip  \noindent {\bf F2: Stateful deception is key for mitigation}  \\ 
We  find static deception strategies can delay but not fundamentally limit the binary success (i.e., infect or exfiltrate critical assets)  of many  attackers we evaluated. In contrast,  stateful deception  that reacts to changes in system state  are able  thwart   many types of attacks we considered (\figRef{fig:eval_goal_heatmap}).

 Essentially,   attackers are quick enough to explore all attack paths even in the face of static deception.
Only the layered deception strategy against a few DFS movement and Persistent attackers are able to stall the attacker to the 75 minute time limit.  In contrast, defenders can use stateful deception strategies to identify and remove attackers from their networks, as seen in \figRef{fig:eval_goal_heatmap}.

\begin{figure}[tb]
  \centering
  % First (top) subfigure
  \begin{subfigure}[b]{\columnwidth}
    \centering
    \includegraphics[width=.95\columnwidth]{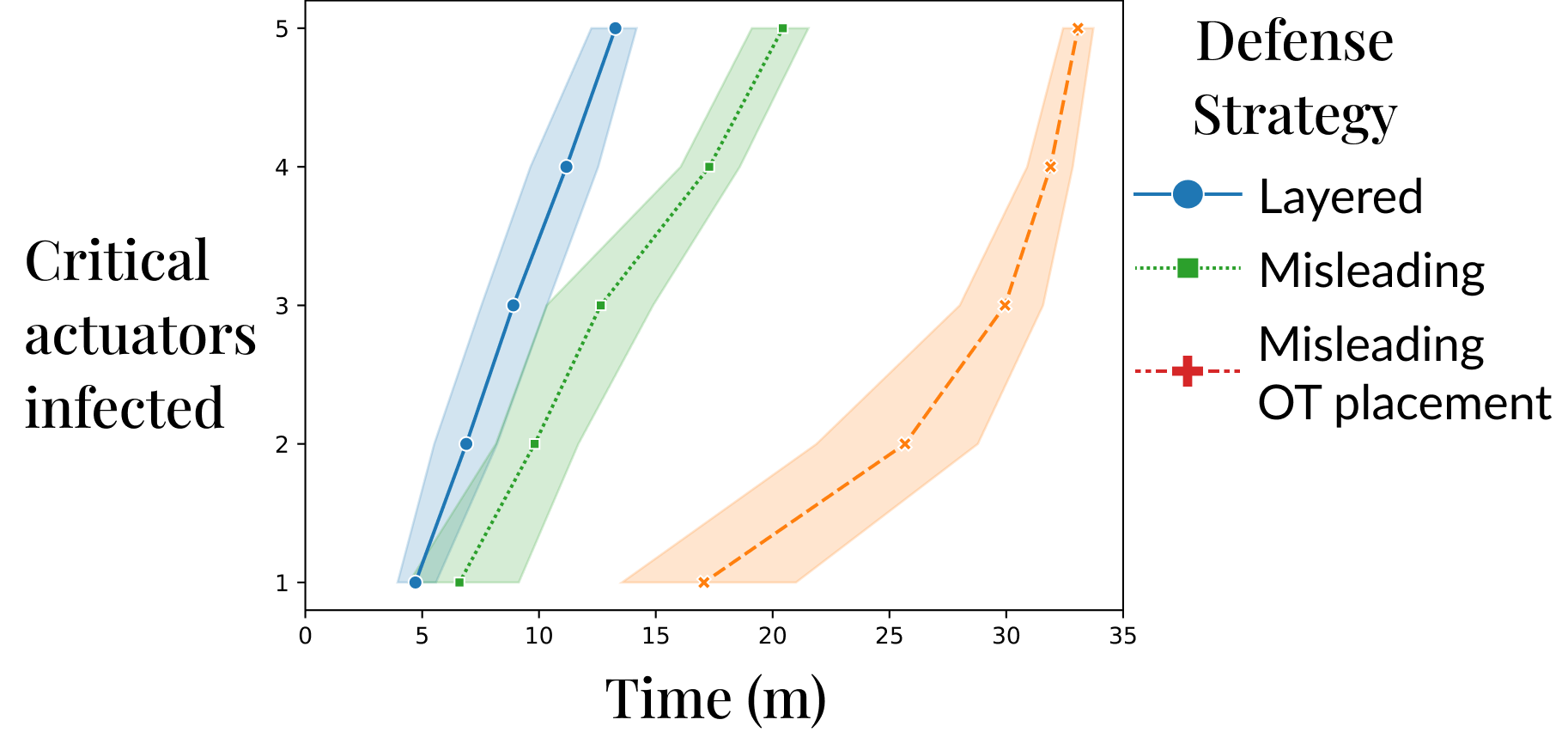}
    \caption{Targeted attacker against three deception defenses in the Colonial Pipeline-inspired environment.}
    \label{fig:eval_targeted_ics_details}
    \label{fig:sys_before}
  \end{subfigure}

  \vspace{1ex} % small vertical gap

  % Second (bottom) subfigure
  \begin{subfigure}[b]{\columnwidth}
    \centering
    \includegraphics[width=.72\columnwidth]{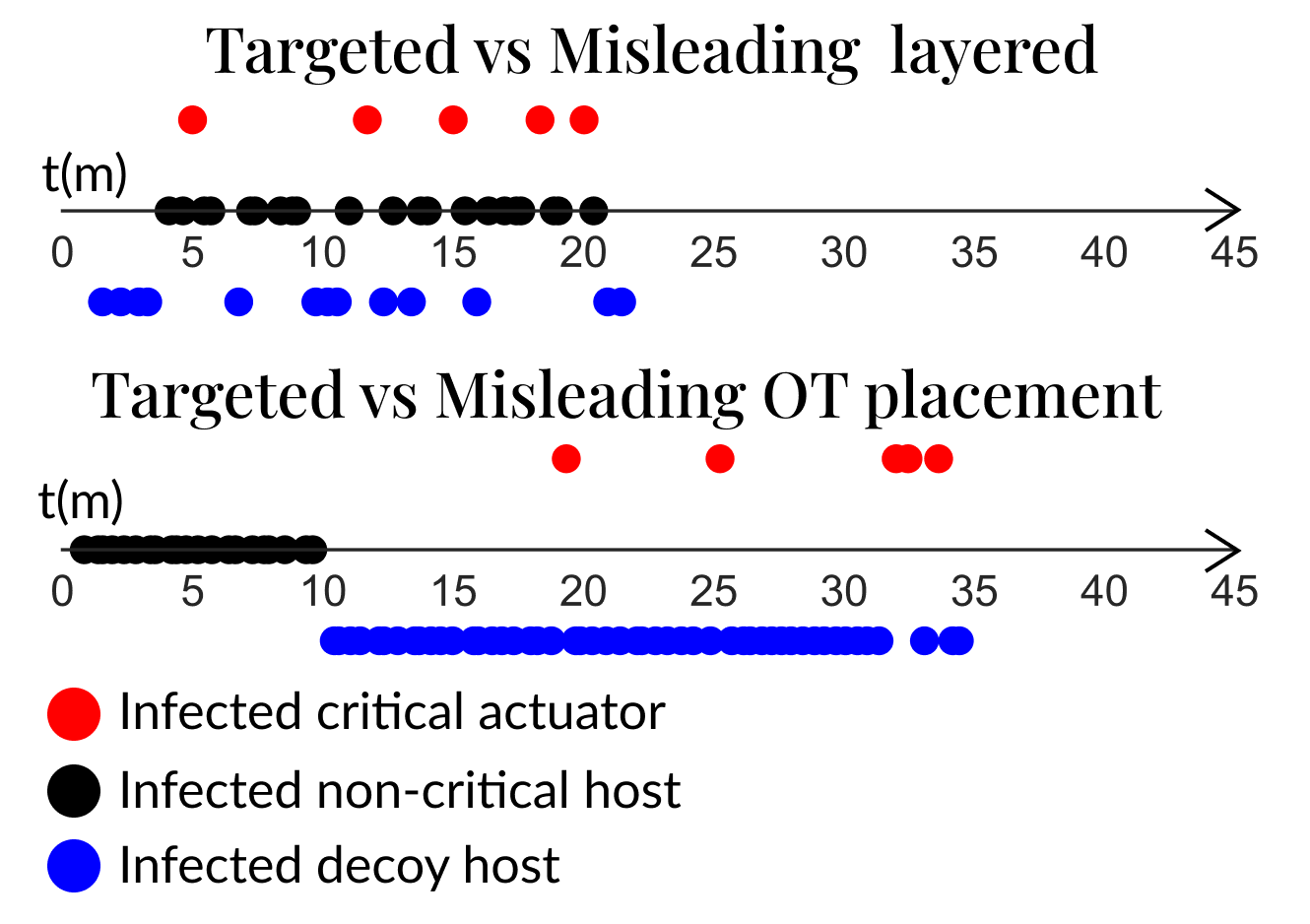}
    \caption{Two example timelines of the targeted attacker against the misleading layered strategy and misleading OT placement strategies.}
    \label{fig:eval_targeted_timeline}
    \label{fig:sys_before}
  \end{subfigure}

  \caption{In the Colonial Pipeline-inspired environment against a targeted attacker, a misleading OT placement strategy can slow the attacker down by 3.2$\times$ more than a layered strategy.}
  \label{fig:eval_targeted_ics}
\end{figure}

\smallskip \noindent {\bf F3: Simple strategy extensions  can yield significant wins} \\ 
Recall from \secRef{sec:eval_perry_extensions}, we considered misinformation and alternative placement extensions to the layered defender. 
We find that misinformation is effective against the targeted attacker (i.e., one that has some topology knowledge) and slows down the attacker by an average of 54.5\% (\figRef{fig:eval_targeted_ics_details}).

We also discover altering the placement policy can further  amplify the efficacy  slow down the targeted  attacker 3.2$\times$.
To identify why this placement policy is so effective, we show two example timelines of attacks in \figRef{fig:eval_targeted_timeline}.
The placement policy ``floods the zone'' of the OT network at the tradeoff of allowing the attacker to gain access to non-critical hosts.

\begin{table}[tb]
\caption{Mean percentage of goals achieved by attacker across five environments against no deception, layered (10 honeypots, 50 decoy credentials), and SysFlow deception strategies (10 honeypots, 50 decoy credentials) for three trials. The SysFlow layered strategy reduces goals achieved by the DFS and Darkside attackers across all five environments.}
  \label{tab:all-strategy-eval}
  \centering
  \setlength{\tabcolsep}{5pt}
  \begin{tabular}{llccccc}
    \toprule
    \multicolumn{2}{c}{} & \multicolumn{5}{c}{Environment} \\
    \cmidrule(lr){3-7}
    Defender & Attacker & Equifax & Col. & Enterprise & Star & Chain \\
    \midrule
    \multirow{2}{*}{No deception} & DFS & 100 & 100 & 66 & 100 & 100 \\
    & Darkside & 100 & 100 & 66 & 33 & 100 \\
    \midrule
    \multirow{2}{*}{Layered} & DFS & 86 & 100 & 66 & 100 & 100 \\
    & Darkside & 99 & 100 & 66 & 32 & 0 \\
    \midrule
    \multirow{2}{*}{SysFlow Layered} & DFS & 3 & 60 & 6 & 4 & 0 \\
    & Darkside & 0 & 40 & 0 & 0 & 0 \\
    \bottomrule
  \end{tabular}
\end{table}

\smallskip \noindent {\bf F4: Fine-grained telemetry can boost deception efficacy} \\
Recall the fine-grained telemetry extensions we discussed earlier in \secRef{sec:eval_perry_extensions}.
We evaluate the layered deception strategy with and without fine-grained telemetry against the DFS and Darkside attackers in all five environments shown in \tableRef{tab:all-strategy-eval}.
Since, both the attackers and defenders are environment agnostic, we are able to execute these experiments without any modifications to the implementations.

We find that in all five environments, against both the DFS and Darkside attackers, adding fine-grained host telemetry to a layered strategy reduces goals achieved by the attacker by 40 to 100\%.
Furthermore, adding fine-grained telemetry is strictly an improvement than the baseline layered strategy.

\smallskip \noindent {\bf F5: LLM-based attackers can be thwarted by deception} \\ 
 Table~\ref{tab:llm_attacker_results} summarizes running our Sonnet 3.7 Thinking LLM-based attacker for various deception scenarios across 10 trials. (Note that this is different from the use of LLM based code generation for defense strategies in \secRef{sec:eval_rapid_exploration}).  The LLM-based attacker successfully exfiltrated data in 2 out of 10 trials  with no deception as shown in \tableRef{tab:llm_attacker_results}. However, against both deception strategies, it failed to exfiltrate data in any trial. We found that  $\approx 90\%$ of commands from the LLM  were indeed ``wasted''  on deceptive resources.
While these are the early days in autonomous LLM-based attackers~\cite{deng2023pentestgpt, singer2025feasibility}, this early result  suggests deception may be  a promising defense against future autonomous LLM-based attackers.

\begin{table}
\caption{Effectiveness of various deception strategies against Claude 3.7 Sonnet attacker model.}
    \label{tab:llm_attacker_results}
    \centering
    \begin{tabular}{>{\raggedright\arraybackslash}p{0.25\linewidth} >{\centering\arraybackslash}p{0.25\linewidth}>{\centering\arraybackslash}p{0.25\linewidth}}
    \toprule
    \textbf{Defense Strategy} & \textbf{Data Exfiltration Success} & \textbf{Decoy Interaction Rate} \\
    \midrule
    No deception & 2/10 & 0\% \\
    Static deception & 0/10 & 92\% \\
    Layered deception & 0/10 & 90\% \\
    Simple stateful & 0/10 & 88\% \\
    \bottomrule
    \end{tabular}
\end{table}

\section{Discussion and Limitations}
\para{Use cases beyond deception}
Perry can potentially enable operators to rapidly explore defense capabilities beyond deception.
The environments and attackers described in \secRef{sec:eval} are not specific to deception.
Furthermore, some of the defense strategies in \secRef{sec:eval} already use non-deceptive defense techniques; other techniques could easily be added.

Perry could also potentially be used to generate and collect realistic synthetic data to train anomaly detection tools.
For example, Perry could run a wide variety of attackers in an environment to train intrusion-detection systems.

\para{Limitations}
What-if scenarios, including for deception, cannot evaluate strategies unless the setting is fully specified.
For Perry, this means being unable to evaluate defenses against unforeseen attackers.
In the real world, attackers can have unexpected strategies or new capabilities, such as a 0-day exploit~\cite{zero_days}.
Defenders are unable to emulate these situations in Perry because they are unaware of these vulnerabilities.

That said, we believe that there is considerable practical value in understanding known attacks.
In \secRef{sec:eval_rapid_exploration} we illustrate how what-if scenarios can provide useful insights for known attackers.
Additionally, after new attackers are discovered they can be quickly added to Perry (see \secRef{sec:eval}).

In both Perry and existing tools~\cite{caldera, elasticsearch, ibmSecurityQRadar, secgen},
defining the environments in full detail, e.g., as precise digital twins of real environments, can be time consuming.
Although Perry supplies abstractions that aid in specifying environments, complex topologies with many hosts whose configurations are different will nevertheless be laborious to specify.
However, as Perry makes it easy to extend or create new environments, we believe that the cost of defining environments will decrease as users share their environments.

Furthermore, a challenge with all emulation tools is the high resource requirements of emulating large networks.
For instance, we estimate that emulating a 10,000-host enterprise organization for the same number of experiments in \secRef{sec:eval_rapid_exploration} would cost \$50,594 on Google Cloud.\footnote{We assume each virtual machine has the same requirements as the experiments in \secRef{sec:eval_rapid_exploration}: 1 CPU, 1GB RAM, and 5GB of disk space. We assume each experiment takes on average 1 hour.}
To help reduce the cost of evaluating deception on large networks, we have preliminarily evaluated whether experimental results obtained on smaller, simplified environments can carry over to larger, more realistic environments; our experiments suggest that results obtained on smaller environments can carry over to larger environments.
As future work, we plan to more thoroughly investigate how to reliably approximate aspects of the environment such as services, vulnerabilities, and human interactions.

\section{Other Related Work}

\para{Abstract game platforms} One low-effort option for implementing deception what-if experiments are game-theory platforms~\cite{milani2020attackGraph, kiekintveld2015honeypotGameTheory, cranford2018stackelberg}.
These platforms do not evaluate deception approaches in real networks—they are abstract games that allow, for example, the attacker and defender to take turns playing actions.
These platforms do not model key aspects of networks, attackers, and defenders making it unclear how their results apply to real deployments.

\para{Unit-test attack emulation tools} Some attack emulation tools  ~\cite{atomic_red_team, uber_metta, sved_tool, infection_monkey} are designed to unit test a series of low-level attacker actions to test defender tools, such as the rules in an intrusion detection system. 
These types of tools are unable to natively emulate the attackers required for deception what-if scenarios because they lack support for controlling and monitoring multiple hosts in an environment.

\para{Other defender emulation tools} Similar to unit-test attacker emulation tools, there are defender emulation tools\cite{fresco, frenetic, caldera} that cannot natively implement types of deception approaches. 
In one case, SDN security orchestration tools~\cite{fresco, frenetic}, which only operate on the network layer and as a result cannot implement deception approaches that operate on hosts, such as decoy credentials or files.
Another example is Caldera's defender plugin~\cite{caldera}, which cannot talk to the network control plane and lacks support for telemetry—two critical components for deception approaches.
For instance, the Caldera defender plugin is unable to deploy decoy hosts because it cannot natively talk to the network control plane.

\section{Conclusions}
Our work was inspired by the significant obstacles we faced in emulating attackers, deception approaches, and environments with existing tools. 
Our experiences and challenges helped us identify common design patterns to inform Perry's programming model for implementing deception what-if experiments.
We introduce Perry to demonstrate the importance of what-if experiments for cyber deception.
Our work can serve as the basis to provide operators with the tools to explore and optimize deception strategies in their deployment.

\textbf{Acknowledgments} This work was sponsored in part by NSF under award CNS2106214.
This work was also supported in part by Microsoft corporation and through the Carnegie Mellon Cylab Future of Enterprise Security initiative.
Some of the initial design was also  sponsored in part by the Combat Capabilities Development Command Army Research Laboratory and was accomplished under Cooperative Agreement Number W911NF-13-2-0045 (ARL Cyber Security CRA). 
The views and conclusions contained in this document are those of the authors and should not be interpreted as representing the official policies, either expressed or implied, of the Combat Capabilities Development Command Army Research Laboratory or the U.S. Government. 
The U.S. Government is authorized to reproduce and distribute reprints for Government purposes not withstanding any copyright notation here on.

\bibliographystyle{plain}
\bibliography{refs}

\begin{appendices}

\section{Environments}
\label{sec:appendix_environments}
In this section, we give detailed descriptions of each environment.

\para{Equifax-inspired environment} The Equifax-inspired environment has two web servers running a vulnerable version of Apache Struts with CVE-2017-5638, the same as the real environment~\cite{equifax_report}.
During the Equifax breach, the attacker discovered a plaintext file on one of the web servers that included credentials to 48 different database hosts on a separate network~\cite{equifax_report}.
\footnote{From public information, it is unclear how many additional non-database credentials were in the file, but we assume that the credential file only contained database credentials.}

To replicate the databases in our environment, we create a second network with 48 database hosts and add files with fake critical consumer data such as emails, social security numbers, and addresses.
On a random web server, we add a plain-text SSH configuration file that contains credentials to all the databases.

\para{Colonial Pipeline-inspired environment}
Next, we implement an environment inspired by the Colonial Pipeline breach~\cite{colonial_pipeline_techtarget} and other ICS attacks~\cite{lee2017crashoverride, sheddingLight}.
The goal of the attacker is to gain access to devices that control physical devices, we call these devices critical actuators.
The environment has three networks: two IT networks and one OT network.
The two IT networks have 10 hosts each and the OT network has 15 sensor hosts, 5 controller hosts and 5 critical actuator hosts.
Each of the 5 controller hosts have credentials to one of the 5 actuators because controller hosts use data from the sensor hosts to control the actuators~\cite{icsControlHosts}.
In addition, each IT network has a management host that have credentials to all sensors and control hosts~\cite{icsControlHosts}.
The monitoring hosts have software that has been compromised by a reverse shell backdoor~\cite{brokenSesameReverseShell}.

Attackers often get access to IT hosts through techniques such as exploiting weak passwords (the case in the Colonial Pipeline breach~\cite{colonial_pipeline_techtarget} and phishing~\cite{smsPhishing, googlePhishingEval}. 
We emulate this in the environment by giving the attacker initial access to a random host on the IT network.

\para{Chain} Some  evaluations of game-theory based deception algorithms consider a Ring  network where each host has credentials to one  other host in the network~\cite{mirage, ringNetwork2013technique}.
Each host has some  critical data and the goal of the attacker is to exfiltrate all the data in the network.
Existing studies simulated the Ring environment with five hosts~\cite{mirage} and 10 hosts~\cite{ringNetwork2013technique}.
We implement our ring network with 25 hosts~\cite{equifax_report}.

\para{Star} Some prior work evaluating deception has also considered a star network~\cite{ferguson2021_deception_psychology}.
The Star network contains 25 hosts on the same network.
The attacker gains access to a host with credentials to all 25 other hosts, creating a star attack graph.

\para{Enterprise} The enterprise network is modeled based on a common tree hierarchy~\cite{ciscoEnterpriseNetwork}.
The Enterprise network has 4 networks, 3 networks represent 3 floors in a building, and the 4th network is for external services.
The external network contains vulnerable webservers.
Each floor has a host vulnerable to a reverse shell backdoor.
One of the vulnerable hosts has credentials to critical databases.
The other vulnerable hosts have credentials to non-critical user servers.

\section{Attackers}
\label{sec:appendix_attackers}

\subsection{Attackers}
Next, we give detailed descriptions of each attacker:

\textit{Equifax attacker} 
We use the  public information  of the breach to define the attacker strategy~\cite{equifax_report}.
The  Equifax attacker is a multistage attacker that 1) finds and infects web servers
to gain initial access, 2) finds and uses credentials found on web
servers to infect internal databases, and 3) finds and exfiltrates
data by copying data onto web servers and downloading it over HTTP.  
The  attacker repeats these steps
until all web servers are infected, all found credentials are used, and all found data is exfiltrated.
Note that the attacker is adaptive to the environment in that it discovers information at runtime. 
The attacker can work in environments with any number of web servers or databases, environments with credentials on different web servers, and data can be exfiltrated from any host (not only hosts on the database network).

\textit{DFS movement}
As a more general attack strategy,  we consider a depth-first-search (DFS) movement attacker to serve as a baseline for the Colonial Pipeline-inspired and Ring environments.
The DFS attacker explores the network with a depth-first search over possible targets.
After infecting a new host, the DFS attacker 1) discovers host
information, 2) tries to exfiltrate any critical data on the host, and
3) adds newly discovered network targets to the start of a stack, and 4) infects the network target at the start of the stack.
For example, if the DFS attacker has three possible web servers to infect, it will infect one web server.
Then, if the attacker finds new credentials, it will exploit these credentials before infecting the other web servers.

\textit{Targeted} The targeted attacker uses the attack graph service to maintain a list of all potential targets in a network.
The targeted attacker prioritizes the targets in the list based on leaked information about the network.
For instance in the Equifax environment, the targeted attacker will prioritize infecting all attack paths that include database servers.

\textit{Persistent} 
The persistent attack strategy reserves an infected host in the network, we call this the persistent host.
If the attacker loses access to a host, the attacker uses the persistent host to reinfect the host they lost access to.
In addition, the persistent attack strategy stores a history of executed actions.
If the persistent attack strategy loses access to a host and then regains access, it will not re-execute the actions on that host.
For example, if a persistent attacker used a decoy credential and lost access to the host, the persistent attacker will reinfect the host, but it will not reuse the decoy credential. 

\textit{Darkside} The Darkside attack strategy is based on the public report of the Darkside attack group~\cite{darkside}.
The strategy first attempts to gain an initial foothold in the network.
Once the attacker gains access it will conduct the following actions repeatedly: escalate privileges, conduct internal reconnaissance, and infect internal hosts.
Once the attack has tried infecting all hosts, they will complete the mission by exfiltrating any found data.

\section{Deception strategies}
\label{sec:appendix_defenders}
Now, we provide detailed descriptions of each deception strategy.

\para{Basic honeypot} The basic honeypot strategy deploys 10 honeypots randomly across each of the defender's networks.
In the Equifax environment, the honeypots are replicas of the vulnerable web servers.
In the other environments, the honey pot is vulnerable to a reverse bash shell.

\para{Mixed deception} The mixed strategy deploys decoy hosts to random subnets and decoy credentials on random hosts.
The decoy credentials are fake, and if the attacker tries using a decoy credential it will not work.

\para{Layered deception} The mixed strategy deploys decoy hosts to random networks and decoy credentials on random hosts. 
In layered strategy, the defender creates real decoy credentials that have permission to a random decoy host.
Furthermore, the defender adds fake data to each decoy host.

\section{Microbenchmarks}
\label{sec:app_micro_benchmark}
\begin{figure}[tb]
    \centering
    \includegraphics[width=0.32\textwidth]{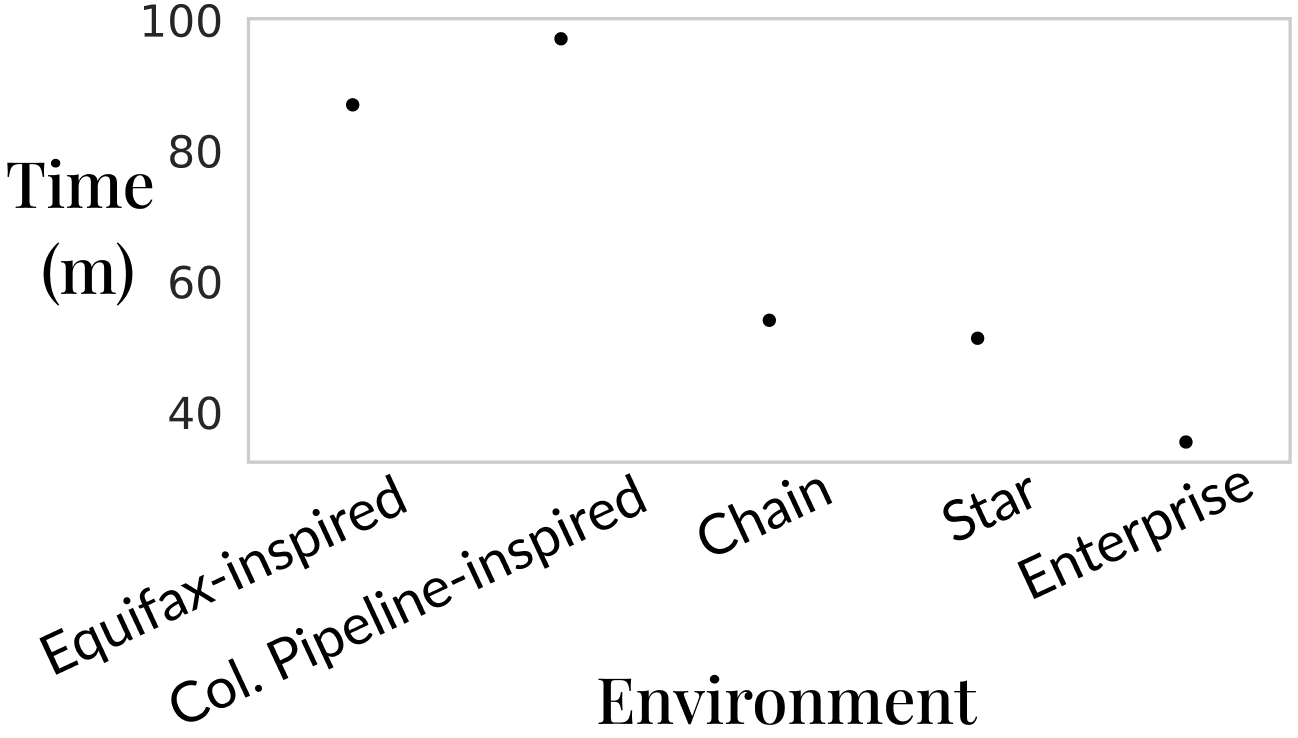}
    \tightcaption{The time taken to setup each environment in Perry. 
    The setup process installs services, vulnerabilities, and other required dependencies.
    In these environments can take from 35–97 min.}
    \label{fig:eval_setup_time}
    \vspace{-0.2cm}
\end{figure}

\begin{figure}[tb]
    \centering
    \includegraphics[width=0.32\textwidth]{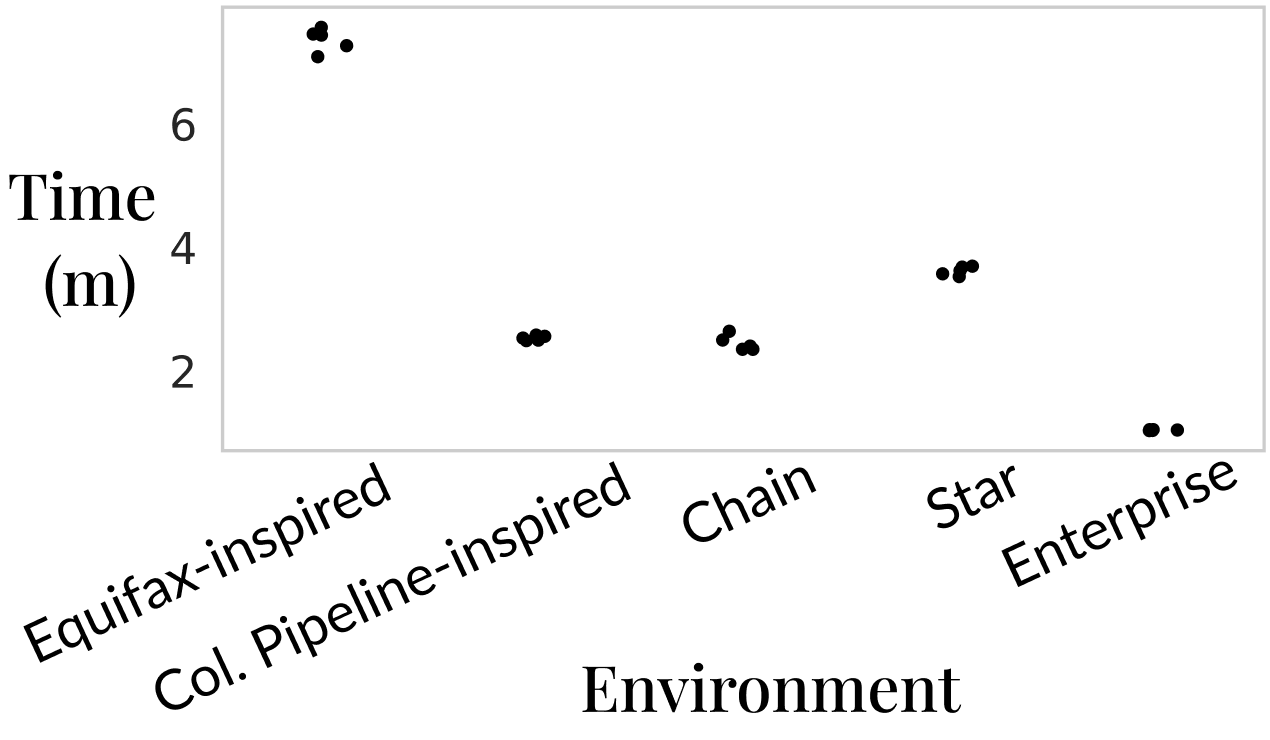}
    \tightcaption{After setting up environments, Perry launches them per scenario.
    We run 5 launches per environment, the launching process ranges from 1–7.3 min.}
    \label{fig:eval_launch_time}
    \vspace{-0.2cm}
\end{figure}

Finally, we benchmark the  time it takes to setup and launch  diverse environments.
As discussed in \secRef{sec:implementation}, Perry instantiates environments in two stages: setup and launch.
For the environments in Table \ref{tab:eval_environments}, the setup times range from 35–97 min., shown in \figRef{fig:eval_setup_time}.
Setup times can be costly because of the cost of downloading and installing required dependencies for tens of hosts in a network. 
After setup, Perry caches the environment and can then relatively quickly launch each environment.
For the environments in Table \ref{tab:eval_environments}, the launch times range from 1–7.3 min. shown in \figRef{fig:eval_launch_time}. 
The key takeaway is that the one-time setup cost is manageable for a new environment and the time to launch is also reasonable to enable the scale of experiments needed. 

\begin{figure}[tb]
    \centering
    \includegraphics[width=0.4\textwidth]{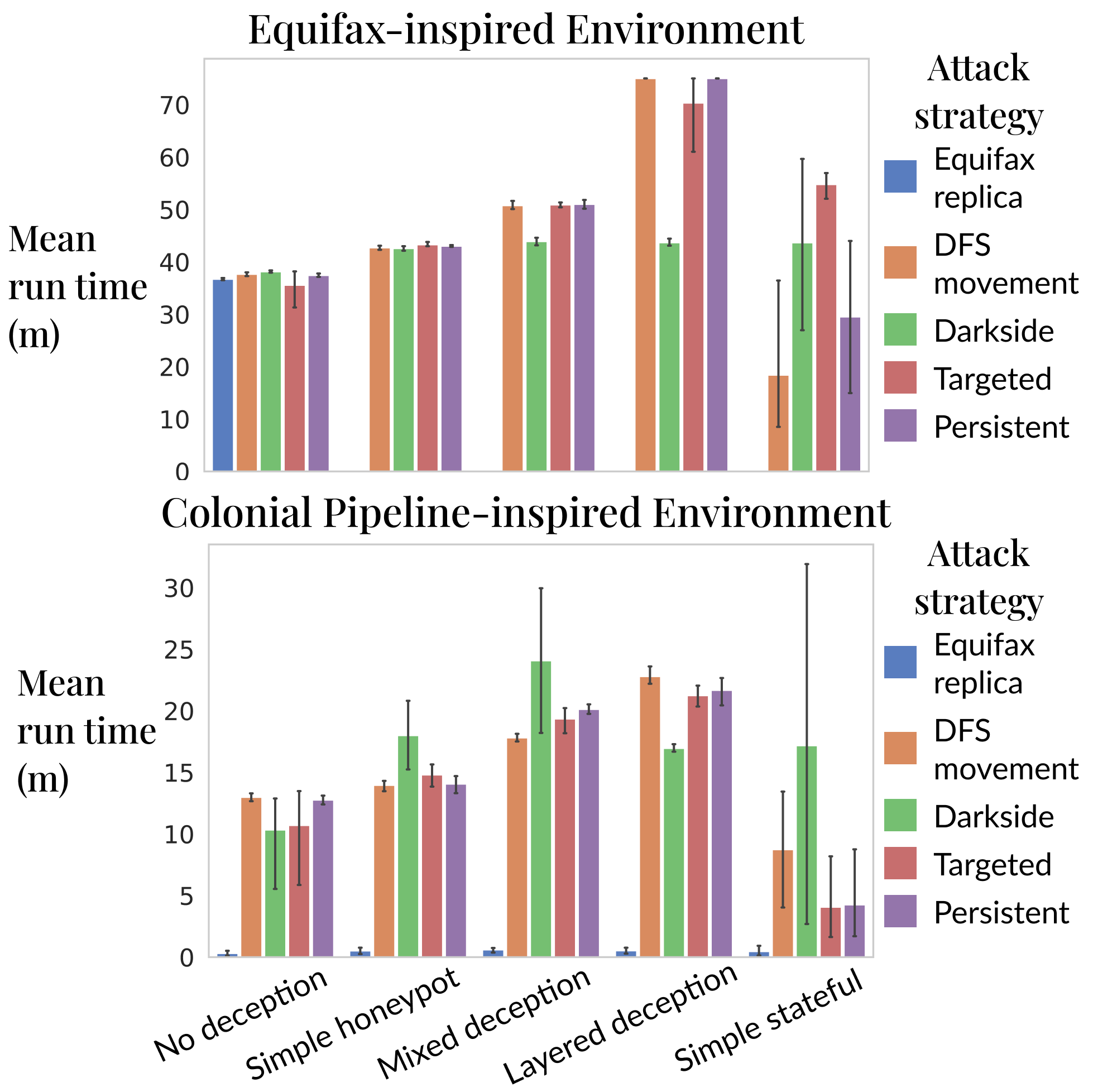}
    \tightcaption{Mean run time of \equifaxEnvShort\ and \cpEnvShort\ experiments in \secRef{sec:eval_rapid_exploration}. The run times range from 0.12–75 minutes.}
    \label{fig:eval_equifax_run_time}
    \vspace{-0.2cm}
\end{figure}

In addition, we measure the run time of each experiment in \secRef{sec:eval_rapid_exploration}.
Experiments have a fairly large range in run time, ranging from 0.12–75 minutes, shown in \figRef{fig:eval_equifax_run_time}. 
Experiments are often short because of attackers failing to make progress infecting the network or being detected by deception.
The longer experiments are from attackers being heavily distracted by deception.

\section{Counting lines of code}
\label{sec:appendix_loc}

In this section, we give a detailed description of how we count and compare lines of code of implementations with and without using Perry's abstractions.

For attackers, we programmatically convert our attacker implementations for Perry into native Caldera code as follows.
First, we preprocess all files to only include semantic lines of code (\ie we remove comments and logging) and we format each file with the PEP 8 format standard. 
We convert Perry attackers by first translating high-level actions into low-level calls to the Caldera SDK.
Next, we convert each call to the telemetry-translation module and environment state service to native Caldera code.
The final result is a functionally equivalent implementation in native Caldera code.

For defenders, we programmatically convert each deception approach into Elasticsearch with native Python code. 
Similarly as for the attacker translation to Caldera,
we convert the high-level actions, telemetry-translation module calls, and environment state service calls into native Python code.
The result is a functionally equivalent implementation using Elasticsearch with native Python code.

\section{LLM generated prompts}
\label{app:llm_prompts}
In \tableRef{tab:llm_translation_result}, we show the specific prompts we queried LLMs for generating deception strategies.

\begin{table}[tb]
  \centering
  \caption{Perry enables LLM-generated deception strategies}
  \label{tab:llm_translation_result}
  \begin{tabular}{p{0.12\linewidth}>{\raggedright\arraybackslash}p{0.5\linewidth}p{0.05\linewidth}>{\centering\arraybackslash}p{0.08\linewidth}}
    \toprule
    \textbf{Deception strategy}  &\textbf{Sonnet 3.7 thinking input}& \textbf{w/o Perry} & \textbf{w/ Perry} \\
    \midrule
    Generic  &Please create a deception strategy.& \emptycirc & $\fullcirc$ \\
    Honeypots  &Please create a deception strategy that deploys 10 honeypots across the network.& \emptycirc & $\fullcirc$ \\
    Advanced  &Please create a deception strategy that deploys a variety of deception resources across the network. If any attackers interact with them, restore the host.& \emptycirc & $\fullcirc$ \\
    \bottomrule
  \end{tabular}
  
  \smallskip
  \begin{tabular}{ll}
    $\emptycirc$ & Generated LLM code does not execute due to errors \\
    % $\halfcirc$ & Code executes \\
    $\fullcirc$ & Generated LLM code executes and is correct \\
  \end{tabular}
\end{table}

\end{appendices}

\end{document}